\documentclass[aps,twocolumn,showpacs,floatfix]{revtex4}

\usepackage{amsfonts}
\usepackage{amsmath}
\usepackage{amssymb}
\usepackage{graphicx}
\usepackage{color}
\usepackage{verbatim}
\usepackage{lipsum}
\usepackage{dsfont}
\usepackage[utf8]{inputenc}
\usepackage{physics}
\usepackage{times}

\usepackage[export]{adjustbox}
\usepackage{dcolumn}
\usepackage{bm}
\usepackage[mathlines]{lineno}

\usepackage[T1]{fontenc}

\usepackage{hyperref}
\usepackage{makecell}
\usepackage{upgreek}
\usepackage{amssymb}
\usepackage{tabularx}

\begin{document}

\preprint{APS/123-QED}

\title{

Impact of helium ion implantation dose and annealing on dense near-surface layers of NV centers

}


\author{A. Berzins$^1$}
\email{andris.berzins@lu.lv}
\author{H. Grube$^1$}
\author{E. Sprugis$^2$}
\author{G. Vaivars$^2$}
\author{I. Fescenko$^1$}

\affiliation{$^1$Laser Center, University of Latvia, Latvia}
\affiliation{$^2$Institute of Solid State Physics, University of Latvia, Latvia}

\date{\today}

\begin{abstract}
Implantation of diamonds with helium ions becomes a common method to create hundreds-nanometers-thick near-surface layers of NV centers for high-sensitivity sensing and imaging applications. However, optimal implantation dose and annealing temperature is still a matter of discussion. In this study, we irradiated HPHT diamonds with an initial nitrogen concentration of 100~ppm using different implantation doses of helium ions to create 200-nm thick NV layers. We compare a previously considered optimal implantation dose of $\sim10^{12}$ to double and triple doses by measuring fluorescence intensity, contrast, and linewidth of magnetic resonances, as well as longitudinal and transversal relaxation times $T_1$ and $T_2$. From these direct measurements we also estimate concentrations of P1 and NV centers. In addition, we compare the three diamond samples that underwent three consequent annealing steps to quantify the impact of processing at 1100~$^{\circ}$C, which follows initial annealing at 800~$^{\circ}$C. By tripling the implantation dose we have increased the magnetic sensitivity of our sensors by $28\pm5$\%. By projecting our results to higher implantation doses we show that a further improvement of up to 70\% may be achieved. At the same time, additional annealing steps at 1100~$^{\circ}$C improve the sensitivity only by 6.6~$\pm$~2.7~\%.

\end{abstract}

\maketitle


\section{Introduction}

The nitrogen-Vacancy (NV) centers in diamond are point defects consisting of a vacancy in the diamond lattice adjacent to a substitutional nitrogen atom~\cite{ashfold_nitrogen_2020}. Negatively charged NV$^-$ centers, which acquire an additional electron mostly from other substitutional nitrogen atoms, possess long coherence times of their electron and nuclear spins and can be initialized and read optically~\cite{barry_sensitivity_2020}. This made them widely studied as potential qubits and quantum sensors. Intensive studies of NV centers in the last decade have led to a large variety of sensing applications~\cite{wu_diamond_2016,chipaux_nanodiamonds_2018,norman_novel_2020,fu_sensitive_2020,abe_tutorial_2018}, which benefit from nanometer resolution and room-temperature operation of the NV-based devices, as well as from low toxicity and mechanical or chemical durability of their diamond matrix. Mostly these applications exploit the high sensitivity of NV centers to magnetic fields via ground state Zeeman effect by using the Optically Detected Magnetic Resonance (ODMR) detection ~\cite{levine_principles_2019, barry_sensitivity_2020, rondin_magnetometry_2014}. 

There are several methods to create NV centers in the diamond. Nitrogen ion implantation is used in crystals with low initial nitrogen concentration~\cite{kehayias_solution_2017, yamamoto_extending_2013, naydenov_enhanced_2010,  rabeau_implantation_2006, meijer_generation_2005}, and the advantage of this method is the control of nitrogen distribution within the diamond, but the disadvantage is the relatively high damage done to the crystal during the implantation, thus introducing undesirable defects and impurities that
might create charge traps, paramagnetic centers and vacancy chains, leading to increased spectral diffusion and degraded spin coherence properties~\cite{fu_conversion_2010,aharonovich_producing_2009,naydenov_increasing_2010}.
In addition, this method, since it is usually applied to diamonds with low initial concentration of nitrogen, suffers from electron donor deficit leading to lower NV$^0$ to NV$^-$ charge-state conversion efficiency~\cite{luo_creation_2022}.
Another widely used method is electron irradiation~\cite{bassett_electrical_2011,mrozek_characterization_2021,bogdanov_investigation_2021,jarmola_longitudinal_2015}, which creates vacancies in crystals with already sufficient nitrogen concentration. Such electron irradiation produces minimum of undesirable defects, but large electron energies required to create vacancies limit the control of the depth. Therefore, this method is good for fabrication of sensors with uniform NV distribution, where the sensing volume matches the volume of the bulk diamond.
And laser writing~\cite{kurita_direct_2021,chen_laser_2017}, where impulse lasers are used to create the vacancies, is not convenient for creation of NV layers over a wide area due to limited optical depth resolution and spatial inhomogeneity as well as due to relatively high optical power required per unit area, leaving it more suited for creation of single NVs or micrometer sized vacancy regions.

In many applications is desirable to keep high spatial resolution by creating well localized NV ensembles~\cite{kurita_direct_2021,bogdanov_investigation_2021,giri_coupled_2018,wolf_subpicotesla_2015}, for example, thin NV layers for magnetic imaging~\cite{levine_principles_2019, fescenko_diamond_2019,berzins_characterization_2021}.
In addition, dense NV ensembles are desirable since the sensitivity scales with square root of the number of NV$^-$ centers. 
To address these needs, helium ion implantation~\cite{huang_diamond_2013,kleinsasser_high_2016} is developed in the recent decade. Irradiation with lightweight helium ions create less damage in the crystal lattice, and at the same time gives good control over the implantation depth. Besides, this method allows to create
high quality imaging sensors from inexpensive synthetic diamonds with high concentration of nitrogen impurities. Diamonds with 100 ppm of nitrogen could potentially lead to high NV$^-$ concentration, if irradiated with high doses of helium ions. However, we expect some NV$^-$ saturation limit primarily due to deficit of electrons (low NV$^-$/NV$^0$ ratio) because of lack of electron donors and competition from other electron acceptors. Such saturation at irradiation doses of $10^{14}$~He$^+/$cm$^2$ is reported in Ref.~\cite{huang_diamond_2013}, but no other systematic studies of helium implantation doses for HPHT diamonds has been reported since then. The NV$^-$ saturation even at lower irradiation doses is reported in nanometeric-thick profiles of NV centers of CVD diamonds~\cite{favaro_de_oliveira_toward_2016}. The recent studies of NV imaging~\cite{fescenko_diamond_2019, berzins_surface_2021} conservatively used $10^{12}$~He$^+/$cm$^2$ irradiation doses, which might be sub-optimal for HPHT diamond applications.

All aforementioned irradiation methods require annealing to promote migration of vacancies to substitutional nitrogen defects, as well as to heal the crystal. However, the optimal annealing conditions is still a cause for the debate.  For example, there is some uncertainty related to the effects that the longer annealing times and higher temperatures brings: on one hand such a treatment reduces the concentration of radiation-induced defects while maximizing the  NV$^-$/NV$^0$ ratio in nitrogen ion implanted samples and increasing the $T_2$ relaxation time~\cite{yamamoto_extending_2013}, but on the other hand in such samples the higher annealing temperatures leads to a rise in the concentration of the H3 center (an emission center formed by a vacancy together with two nitrogen atoms (NVN))~\cite{sumikura_highly_2020}, that might lead to adverse effects on P1 to NV$^-$ conversion efficiency. In general, existing experimental studies of annealing are hardly comparable, as they are performed using different NV preparation methods and diamonds, at the same time very different annealing procedures are reported in the case studies. There is a body of publications using annealing in temperature interval 750~$^{\circ}$C to 900~$^{\circ}$C and annealing times from 1 to 2 hours \cite{huang_diamond_2013,kleinsasser_high_2016,havlik_boosting_2013,mccloskey_helium_2014} under vacuum or Ar and H$_2$ mixture. 
Some researches apply longer annealing times \cite{tallaire_synthesis_2019,dolde_room-temperature_2013} and higher temperatures \cite{acosta_diamonds_2009,favaro_de_oliveira_toward_2016,sumikura_highly_2020,yamamoto_extending_2013} or both~\cite{fescenko_diamond_2019,bogdanov_investigation_2021,naydenov_increasing_2010}. It is likely that in many cases the temperature range 750~$^{\circ}$C to 900~$^{\circ}$C is defined by maximum temperature achievable by majority of conventional ovens. Besides, additional annealing in air at temperatures around 500~$^{\circ}$C is sometimes used to improve luminescence of NV centers~\cite{havlik_boosting_2013,himics_effective_2015}, but such a treatment is off topic of our study.

In this research we set out to find trends of fluorescence intensities, contrast and FWHM of ODMRs, as well as $T_1$ and $T_2$ relaxation times for three HPHT diamond samples with a nitrogen concentration of $\sim100$~ppm, which we irradiated with standard (previously used~\cite{fescenko_diamond_2019,berzins_surface_2021}), double and triple $^4$He$^+$ doses to create $\sim200$~nm thick NV layers. We hypothesized that by doubling or tripling the He$^+$ implantation dose of a HPHT diamond would proportionally increase the concentration of NV$^-$ centers, and therefore could lead to fabrication of imaging sensors with higher magnetic sensitivity. We also investigate changes of these parameters after applying each of three consecutive annealing steps: first at maximum temperatures of 800~$^{\circ}$C and two successive annealing at maximum temperature of 1100~$^{\circ}$C.

\section{Experimental methods}

\subsection{Fabrication}

\begin{figure}
      \begin{center}
   \includegraphics[width=0.4\textwidth,valign=t]{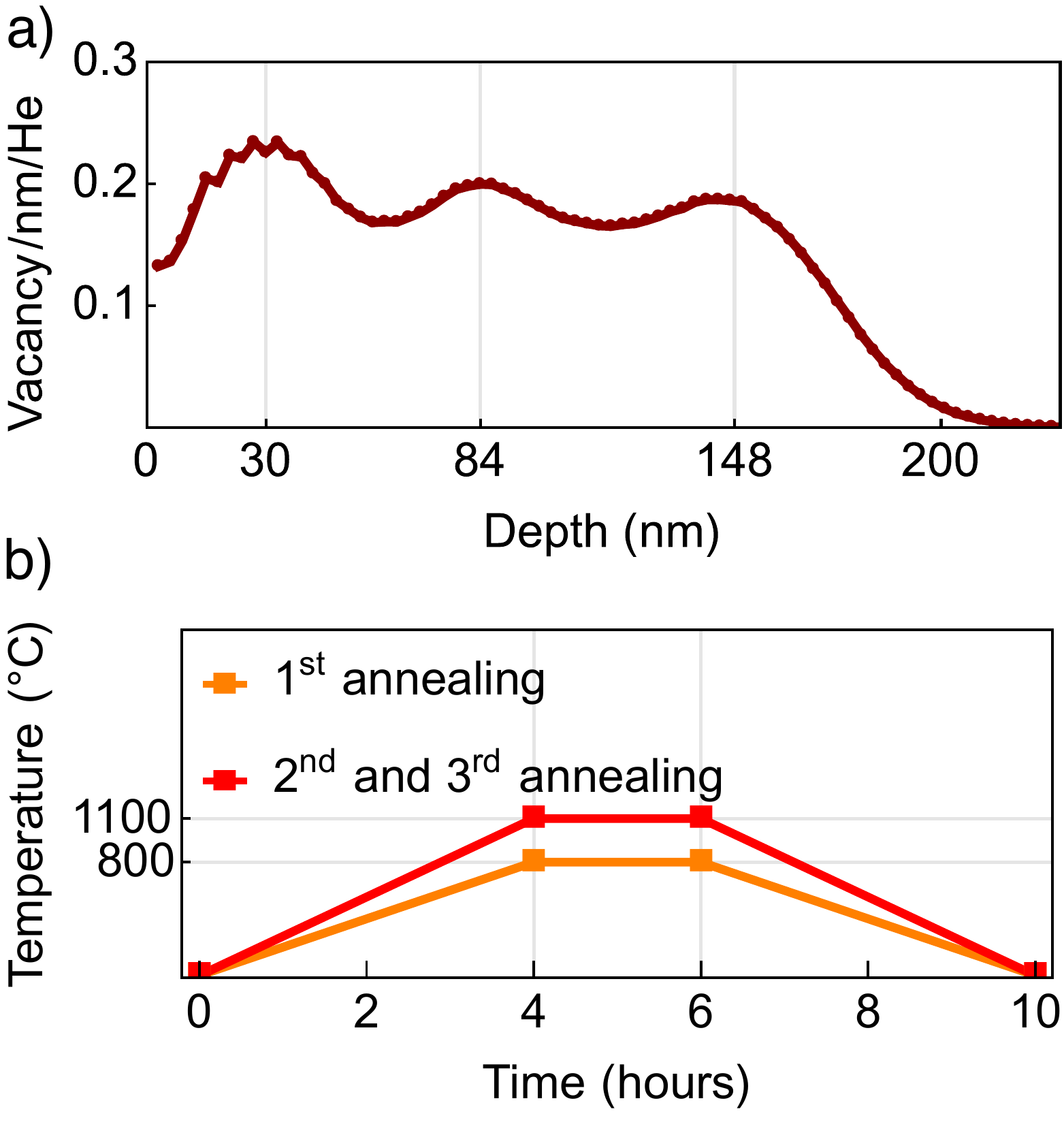}
          \end{center}
       \caption{\textbf{Fabrication of samples: a)} SRIM vacancy-depth profile for for fabrication of 200-nm-thick NV layer close to the diamond surface. \textbf{b)} Time-temperature graphs for the three annealing steps.}
  \label{fabrication}
\end{figure}

In measurements we use three HPHT type Ib diamond crystals (Sumitomo Electric) with a (110) surface polish and with dimensions of 2~mm~$\times$~2~mm~$\times$~0.06 mm. All three crystals (samples F1, F2, and F3) are initially cut from one 0.5~mm thick crystal by Almax easyLab BVBA. We performed \textbf{S}topping \textbf{R}ange of \textbf{I}ons in \textbf{M}atter or SRIM simulations~\cite{ziegler_srim_2010} to determine the implantation parameters required for fabrication of 200-nm-thick NV layer close to the diamond surface (FIG~\ref{fabrication} a).

The three crystals are irradiated with He ions at three separate energies 33 keV, 15 keV and 5 keV with doses represented in Table \ref{energy-dose} by Ion Beam Services SA. After the implantation the crystals went through three steps of annealing with 6~h boiling at 200~C$^{\circ}$ in triacid (1:1:1 mixture of nitric:perchloric:sulfuric acids) before and after each step.  The first annealing is done at 800~$^{\circ}$C for two hours, and the last two annealing steps are done at 1100~$^{\circ}$C (FIG~\ref{fabrication}b). All annealing steps are done under vacuum and in all cases the ramp up time and cool-down time is 4 hours. First annealing step is done by using a Setaram LABSYS evo STA system and in 1$\cdot{10^{-2}}$ $\pm$~0.1$\cdot{10^{-2}}$~mbar vacuum, but the last two annealing steps were done in tube furnace (OTF-1200X-S from MTI corporation) in 1$\cdot{10^{-5}}$ $\pm$~0.3$\cdot{10^{-5}}$~mbar vacuum (Edwards T-Station 85H Wet).
After each annealing step a full set of measurements is performed for each of samples in six equidistant spots along a diagonal of the sensor's top surface.
We take a mean value of the all measurements in the six spots with its standard error as an error bar.  

\begin{table}[h]
\begin{tabularx}{0.4\textwidth} { 
   >{\centering\arraybackslash}X 
   >{\centering\arraybackslash}X 
   >{\centering\arraybackslash}X 
   >{\centering\arraybackslash}X 
   >{\centering\arraybackslash}X }
 \hline
 \hline\\
 Energy, keV & Normalised dose & \multicolumn{3}{c}{Dose ($10^{12}$ He$^+$/cm$^2$) } \\
 &&F1 & F2 & F3 \\
 \hline
 33 & 1.0 & 4.0 & 8.0 & 12.0 \\
 15 & 0.5 & 2.0 & 4.0 & 6.0  \\
 5 & 0.5 & 2.0 & 4.0 & 6.0  \\
\hline
  & Total: & 8 & 16 & 24  \\
\hline
\hline
\end{tabularx}
\caption{He ion implantation doses and energies used for fabrication of samples F1, F2, and F3.}
\label{energy-dose}
\end{table}

\subsection{ODMR measurements}
Firstly, we characterize samples by measuring and analysing their CW ODMR spectra~\cite{levine_principles_2019, barry_sensitivity_2020}. Zeeman splitting between ground-state electronic spin levels is induced in the NV$^-$ centers by a bias magnetic field applied along one of four possible NV axes. We detect a fluorescence spectrum containing a series of separated magnetic resonances by sweeping a transverse to the NV axis microwave field. To quantify ODMR contrast, full width at half maximum (FWHM), and fluorescence intensity off resonance we fit the spectrum with a series of Lorentzians. 
Both contrast and FWHM are obtained from the fit of a resonance at spin transition $|0\rangle\leftrightarrow|-1\rangle$. Measuring the FWHM linewidths we keep the MW power weak enough to avoid any power broadening (see an inset in FIG.~\ref{Fluorescence} c)).

The off-resonance fluorescence intensity gives information about NV=NV$^-$+NV$^0$ concentration in the samples. Other fluorescent centers that contribute to the fluorescence, like H3 center (NVN)~\cite{sumikura_highly_2020} or helium vacancies (HeV)~\cite{forneris_creation_2016,tallaire_synthesis_2019} are much less abundant or not radiate in the detection frequency range. 
The contrast (the relative fluorescence intensity difference in ODMR signal on and off resonance) provides further information about charge of NV centers as it is proportional to NV$^-$/(NV$^-$+NV$^0$) ratio. The FWHM informs about inhomogeneity of NV environment that represents a limiting factor for the magnetic field sensitivity of CW ODMR methods. This FWHM is directly related to inhomogeneously broadened transverse relaxation time $T_2^*$ and is caused by several NV spin ensemble dephasing sources, like interactions with nuclear $^{13}$C bath spins~\cite{bauch_ultralong_2018,dreau_high-resolution_2012,mizuochi_coherence_2009}, crystal-lattice strain fields over the diamond~\cite{bauch_ultralong_2018,jamonneau_competition_2016}, and measurement-related artifacts such as magnetic field gradients over the collection volume and temperature fluctuations~\cite{bauch_ultralong_2018,acosta_temperature_2010}.

\subsection{Relaxation measurements}

Secondly, we characterize samples by measuring and analysing dynamics of NV ensembles by using relaxometry measurements: longitudinal (spin-lattice) relaxation time $T_1$ that characterizes NV spin ensemble dephasing mainly due to cross-relaxation within the strongly interacting bath of NV$^-$ spins~\cite{bauch_decoherence_2020,jarmola_temperature-_2012}; and the transverse relaxation time $T_2$ that characterizes homogeneous decoherence of the prepared state of the NV ensemble, and are mainly caused by interaction of NV$^-$ with spin bath of  substitutional nitrogen atoms (P1 centers)~\cite{bauch_decoherence_2020}.
For detailed description and explanation of these relaxometry measurement sequences see references~\cite{levine_principles_2019, barry_sensitivity_2020, rondin_magnetometry_2014}.

The used microwave sequences are preceded by a $5~\upmu$s long initializing laser pulse to prepare the population in the $|0\rangle$ ground state. For $T_1$ sequence we use a $\{(\pi)-\tau\}$ and for $T_2$ (Hahn echo) we use $\{\pi/2-\tau/2-\pi-\tau/2-\pi/2\}$ MW impulse sequences, where $\tau$ is interrogation time, $\pi$ denote microwave pulse that transfers NV$^-$ population between ground-state electronic spin levels, but $\pi/2$ microwave pulse creates a superposition of these levels. We start every second run of the $T_1$ sequence with a $\pi$ pulse in order to alternate interrogation of population on $|0\rangle$ and $|+1\rangle$ spin levels. The same alternation for the Hahn echo sequence is done by shifting a phase of the last $\pi/2$ pulse relative to the first $\pi/2$ pulse by $90^{\circ}$ in every second run. The $5~\upmu$s long read-out laser pulse induces a fluorescence pulse of a similar shape to the initializing pulse at the start of the sequence, but with a signal depression in the beginning. This relative amplitude of the signal is proportional to a population of the interrogated level. From the difference between the the fluorescence signals of the initializing pulse and read-out pulse we calculate a common-noise-free ODMR contrast, which is plotted as a function of the  increasing interrogation time $\tau$. The resulting decay plots are fitted with exponential functions in the form $C$~exp$(-\tau/T)^p$ where $C$ is contrast, $T$ is a relaxation constant, but parameter $p$ is 1 for fitting longitudinal decays or 3/2 for fitting transverse decays~\cite{bauch_decoherence_2020}.

\subsection{Apparatus}

\begin{figure}
      \includegraphics[width=0.47\textwidth]{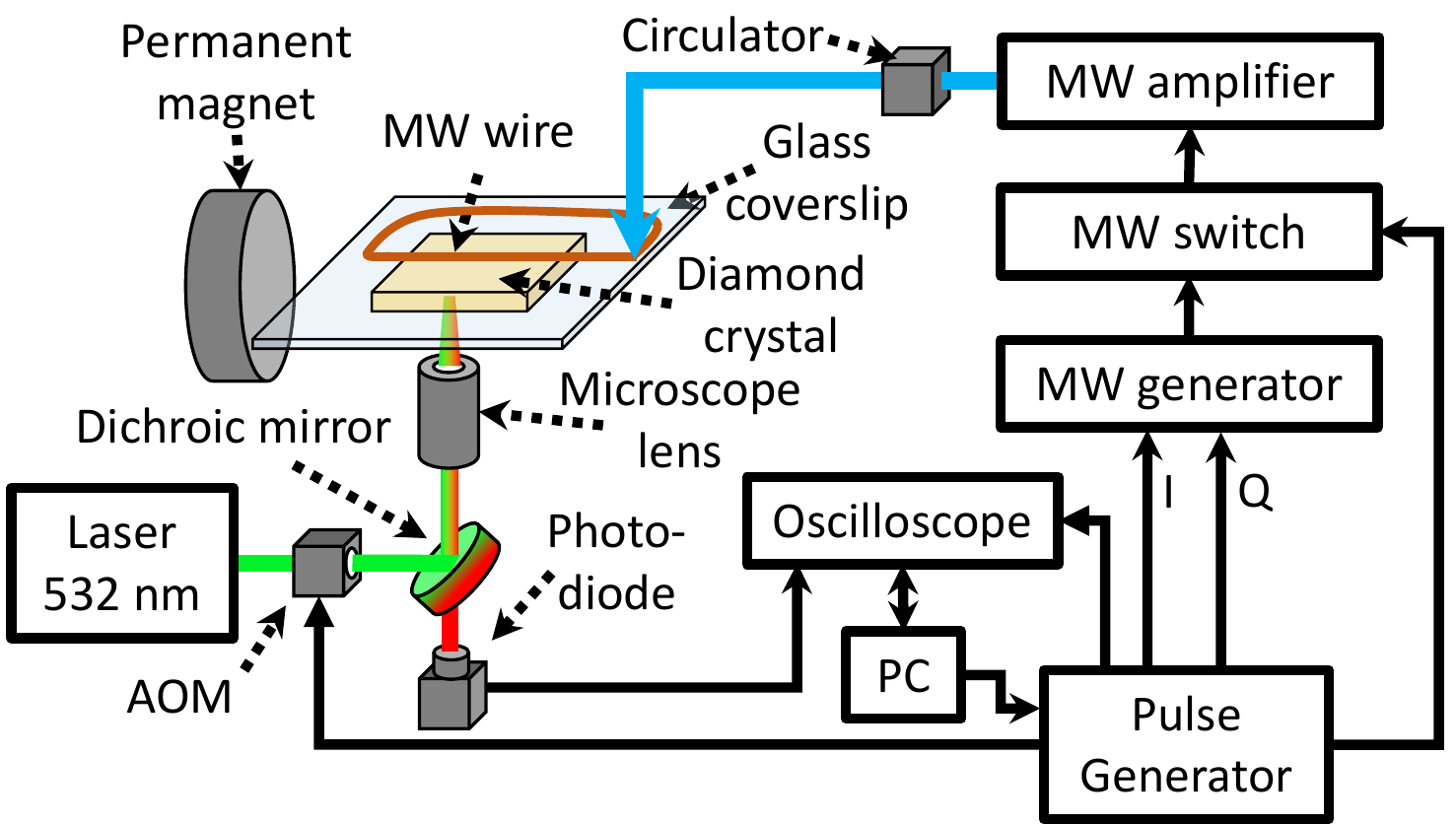}
    \caption{\textbf{Schematic of the experimental apparatus.}  AOM: acousto-optic modulator; I and Q: phase shift control; PC: personal computer.}
  \label{setup}
\end{figure}

The experimental setup for characterization of samples is depicted in Figure~\ref{setup}.
During the measurements the diamond sample is placed on a coverslip in an epifluorescent microscope, in which the NV excitation and fluorescence detection are performed through the same oil-immersion infinity-corrected 100$\times$ microscope objective with numerical aperture of 1.25 (ZEISS). The NV centers are exposed to 200~mW radiation guided by a multi-mode optical fiber and lens system from a Coherent Verdi V-18 laser. The NV fluorescence ($650\mbox{--}800$ nm) is separated by a dichroic mirror (Thorlabs DMLP567R) and is measured on an avalanche detector (Thorlabs APD410A/M) through a long-pass filter (Thorlabs FEL0600). During the measurements we illuminate the NV layer in a region with diameter of 30$~\upmu$m.
 
The bias magnetic field $B_0\approx6$~mT is produced by a neodymium permanent disk magnet and aligned along one of the NV axes in the plane of the diamond plates (polished along the (110) direction).
The MW field used for the measurements is produced by a microwave generator (SRS SG384). The microwaves subsequently pass through an amplifier (Mini-Circuits ZHL-16W-43+) and circulator and are delivered by a copper wire with diameter of 50$~\upmu$m to the diamond sensor. 

The relaxation measurements are controlled by a TTL pulse card (PBESR-PRO-500 by SpinCore). Microwave pulses are generated using the microwave generator in the I/Q modulation mode. The microwave amplitude and phase are controlled on a $\lesssim10$ ns timescale using a series of TTL controlled switches (Mini-Circuits ZASWA-2-50DR). Laser pulses are generated by passing the continuous-wave laser beam through an acousto-optic modulator (MT200-A0,5-VIS by AA Optoelectronic).
An oscilloscope measures the avalanche detector output voltage, reporting fluorescence time traces to the computer controlling the experiment. 

\begin{figure*}
\begin{center}
\includegraphics[width=0.90\textwidth,valign=t]{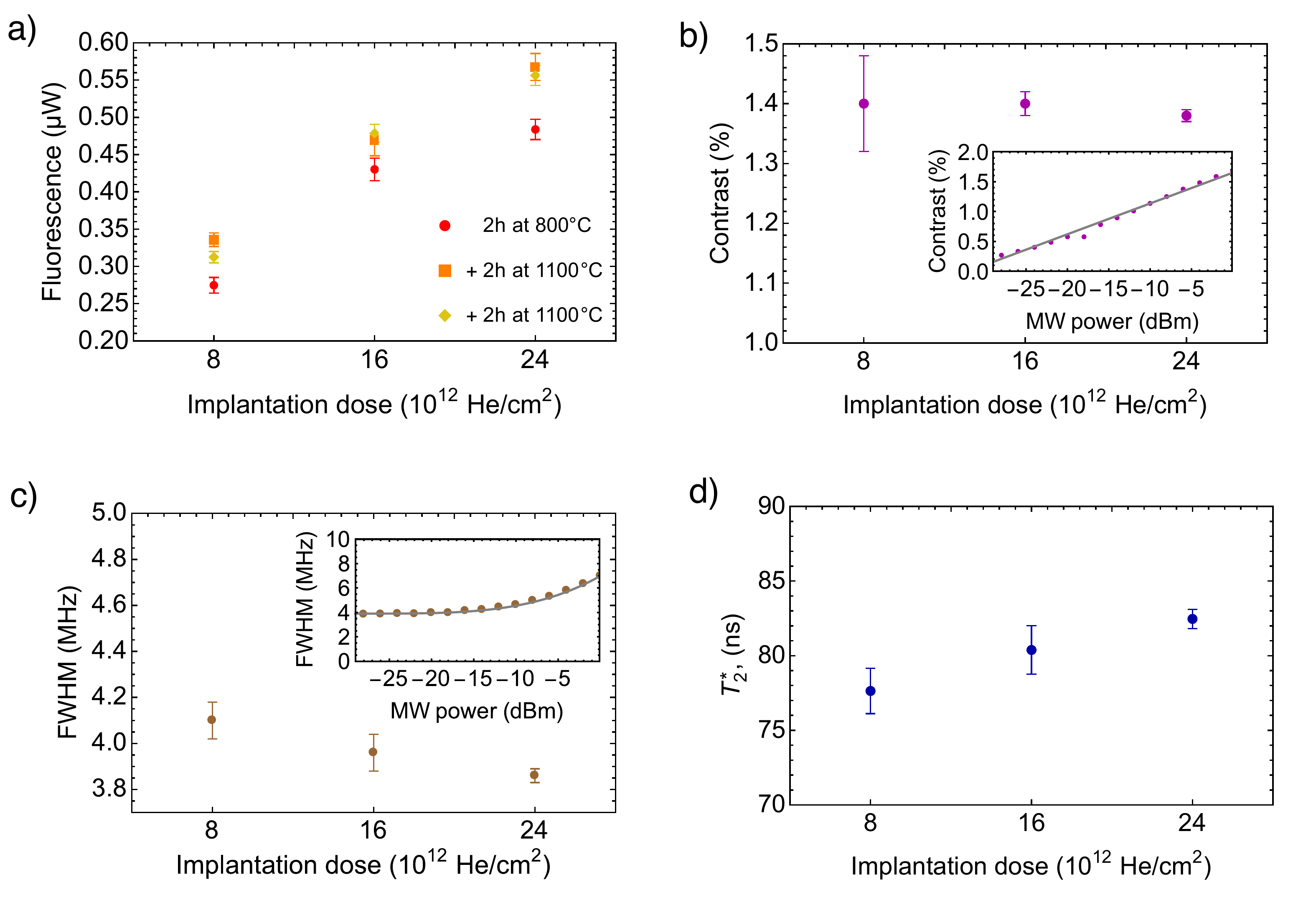}
\hspace{0.1cm}
\end{center}
\caption{\textbf{ODMR measurements: a)} Off-resonance intensity of fluorescence for three subsequent annealing steps versus cumulative implantation dose. \textbf{b)} Contrasts of the ODMR signals at $|0\rangle\leftrightarrow|-1\rangle$ spin transition versus the cumulative implantation dose. The contrasts are shown for $-5$~dBm of MW power (see the inset).  \textbf{c)} The FWHM of the ODMR signals at $|0\rangle\leftrightarrow|-1\rangle$ spin transition versus the cumulative implantation dose. The FWHMs linewidths are shown for $-25$~dBm of MW power (see the inset). \textbf{d)} Calculated from the FWHM inhomogeneously broadened transverse relaxation $T_2^*$ versus the cumulative implantation dose. Complete brakedown of implantation doses  can be found in Table~\ref{energy-dose}. The error bars represent standard error (SE) of the data. All ODMR data except the fluorescence intensity show no significant correlation with the annealing temperature or duration, therefore the contrasts, FWHMs and  $T_2^*$ after averaging over all three annealing steps are shown for simplicity.  
}
\label{Fluorescence}
\end{figure*}

\section{Results and analysis}

\subsection{Fluorescence intensity, contrast, linewidth and $T_2^*$}

The results of ODMR measurements are summarized in FIG.~\ref{Fluorescence}. 
As expected the fluorescence intensity (FIG. \ref{Fluorescence} a)) is larger when larger He$^+$ implantation doses is used, because higher vacancy concentration leads to higher proportion of the P1 centers converted to NV centers.
There is also pronounced fluorescence intensity increase between the first annealing at 800~$^{\circ}$C and the second annealing at 1100~$^{\circ}$C, and an additional third annealing at the last temperature does not lead to a prominent increase of intensity anymore.  That rather means that free vacancies do not travel fast enough to create all potential NV centers during the first annealing, and likely a longer annealing at the same lower temperature would produce the same increase of intensity. In other words, besides the healing the lattice, the annealing at 1100~$^{\circ}$C accomplished the started work of the previous annealing at lower temperature by moving on free vacancy toward substitutional nitrogen atoms. However, previous research~\cite{yamamoto_extending_2013} shows that increasing the annealing temperature to $\approx$1100~$^{\circ}$C enhances the $T_2$ relaxation time (discussed in the next section). An evident question then arises if just a one annealing at 1100~$^{\circ}$C would be enough instead of the more complicated two-step annealing.

The ODMR contrast versus the implantation dose presented in FIG.~\ref{Fluorescence} b) shows no change within error-bars. It also does not have a significant correlation with the annealing temperature or duration, and for simplicity it is shown here after averaging over all three annealing steps. 
The contrast is proportional to the ratio NV$^-/$NV$^0$, as only the negative NV centers contribute to the ODMR signal, but fluorescence from the neutral NV centers contributes to the signal background alone. Such a ratio could drop when most of the P1 centers are converted to NV centers~\cite{huang_diamond_2013,favaro_de_oliveira_toward_2016} because P1 centers are main donors of the electrons for the negative NV centers. The higher implantation dose, the higher concentration of vacancies,  which leads to higher proportion of the P1 centers (single nitrogen defects) converted to NV centers. As a rule of thumb, the concentration of the NV centers should not be larger than the concentration of P1 donors, because a further increase of the NV concentration would not lead to creation of new negative NV centers. The "standard" implantation dose of $10^{12}$~He$^+/$m$^2$ previously used in Ref~\cite{fescenko_diamond_2019,berzins_surface_2021} was chosen because of an estimate that it leads to creation of $\approx50$ ppm of NV centers. In fact, not all population of NV centers acquires the negative charge regardless the abundance of P1 centers, and usually less than 30\% of an ensembles of NV centers is negatively charged~\cite{farfurnik_enhanced_2017,acosta_diamonds_2009}. This indicates that only the comparison between NV$^-$ and P1 center concentrations really matters for determination of NV$^-$ saturation concentration. And likely this is the reason why we do not see a saturation in the NV$^-$ concentration (drop in the contrast) even after the triple dose.

\begin{figure*}
             \includegraphics[width=0.9\textwidth,valign=t]{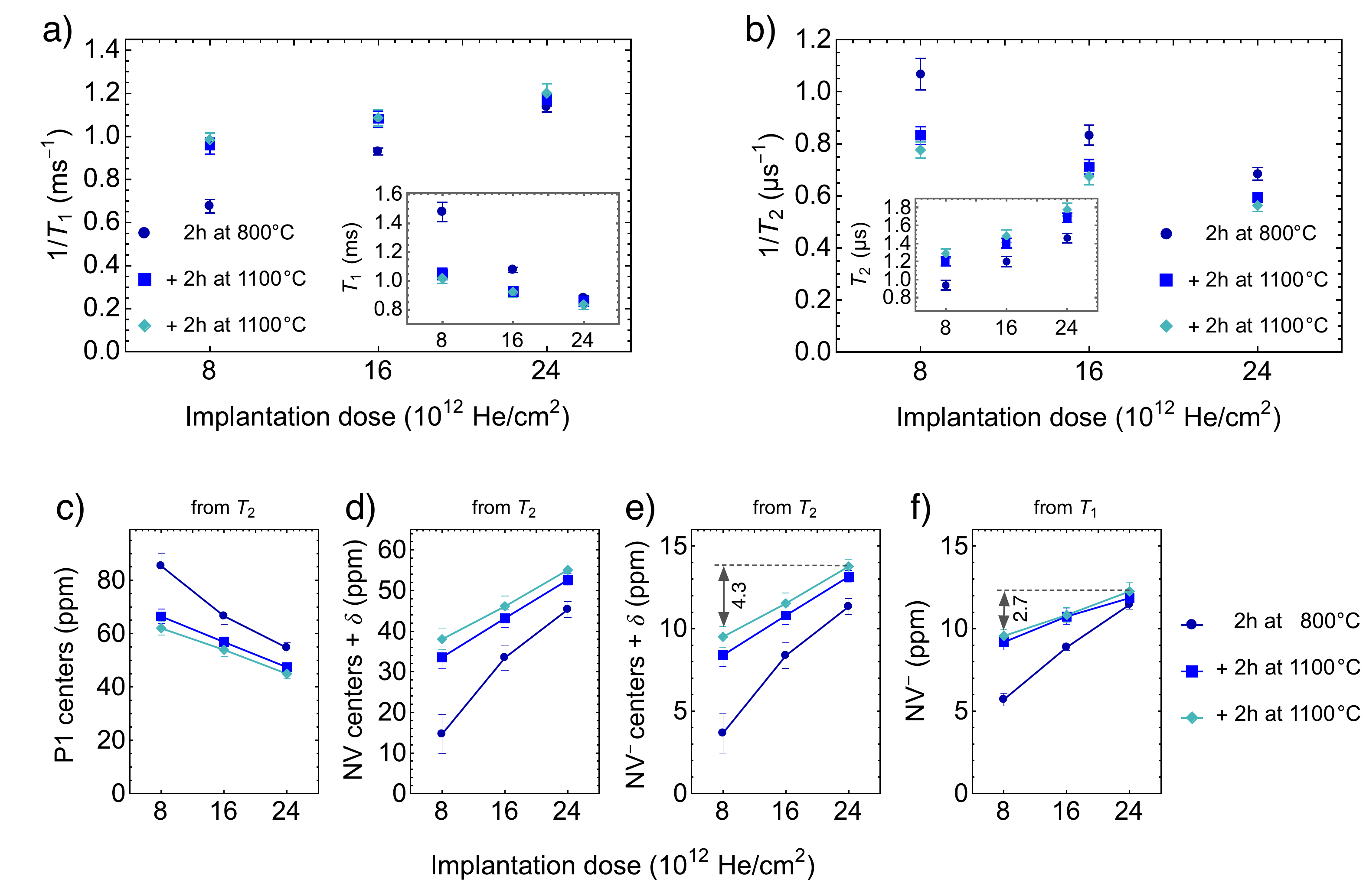}
             \caption{\textbf{Relaxation measurements: a)} Relaxation rate 1/$T_1$ for three subsequent annealing steps versus cumulative implantation dose. The inset shows the same plot in $T_1$ units. \textbf{b)} Relaxation rate 1/$T_2$ for three subsequent annealing steps versus cumulative implantation dose. The inset shows the same plot in $T_2$ units. \textbf{Estimates of concentrations: c)} Concentration of P1 centers estimated from $T_2$. \textbf{d)} Concentration of NV centers estimated from $T_2$. \textbf{e)}  Concentration of NV$^-$ estimated from $T_2$. Symbol $\delta$ denotes small concentrations of other N-containing defects. \textbf{f)} NV$^-$ concentration estimated from $T_1$.
             All estimates are shown versus cumulative implantation dose for three subsequent annealing steps. The solid lines are shown to guide the eye.
        }
  \label{Relaxations}
\end{figure*}

The FWHM linewidth and associated with it relaxation time $T_2^*$ versus the implantation dose are presented in FIG.~\ref{Fluorescence} c) and d), correspondingly. The FWHM does not have a significant correlation with the annealing temperature or duration, and for simplicity it is also shown after averaging over all three annealing steps. The relaxation time calculated from linewidth $\Gamma$  as $T_2^*=1/(\pi\Gamma)$~\cite{ishikawa_optical_2012,acosta_diamonds_2009} is sensitive to magnetic noise of various origin. Because of a decrease in concentration of magnetically noisy P1 centers due to their combination with free vacancies we expect the mitigation of the NV spin dephasing~\cite{bauch_ultralong_2018} and larger values of $T_2^*$ when higher implantation doses are used. Besides, the P1 centers could be converted into H3 centers or NVN, which are not detected by the experimental setup since they radiate at 505.8~nm~\cite{sumikura_highly_2020}. However, the concentration of H3 centers is by two order less than the concentration of P1 centers~\cite{sumikura_highly_2020, deak_formation_2014}, so their contribution to the dephasing is relatively small. The $T_2^*$ plot on FIG.~\ref{Fluorescence} d) qualitatively support the dominant role of P1 centers in the dephasing of NV spins.    

Values of the fluorescence intensity $I$, the contrast $C$ (the relative difference in ODMR signal on/off resonance), and the FWHM linewidth $\Gamma$ allow us to compare the sensitivity of the samples that is the minimum detectable magnetic field of a Lorentzian ODMR signal as~\cite{fescenko_diamond_2019}
\begin{equation}
B_{\rm min}\propto\frac{\Gamma}
{C\sqrt{I}}.
 \label{eq:sen}
\end{equation}
By normalizing the sensitivities obtained with Eq.~\ref{eq:sen} to $B_{\rm min}$ of the sample F1 with the smallest "standard" implantation dose we found the relative improvement of the sensitivity for the sample F2 by $22~\pm~5~\%$ and for the sample F3 by $28~\pm~5~\%$. At the same time, an average improvement of the sensitivity of all samples between the first annealing at 800~$^{\circ}$C and the second annealing at 1100~$^{\circ}$C is the modest 6.6~$\pm$~2.7~\%.

\subsection{Longitudinal and transverse relaxations}

Measured longitudinal relaxation rates versus  cumulative implantation doses are plotted in FIG.~\ref{Relaxations} a). The inset shows the same plot in $T_1$ units. This relaxation characterizes the rate with which the spin population decays back to a thermally mixed state mainly due to cross-relaxation interactions with a bath of other NV$^{-}$ centers~\cite{bauch_decoherence_2020}. The density of NV$^{-}$ bath in our samples varies with concentration of vacancies (implantation dose) and a completeness of the annealing procedure. As a result, we see an increase of $1/T_1$ rate both due to higher implantation doses and partially due to the second annealing step. In perspective of $T_1$, the effect of the second annealing is due to the shortness of the first annealing step, rather than due to larger annealing temperature. This is especially clear for the sample with the largest implantation dose, which is not improved with the second annealing step. Indeed, when a large implantation dose leads to a dense network of the vacancies, a vacancy needs a shorter time to travel before it combines with a nitrogen atom.

FIG.~\ref{Relaxations}~b) depicts measured $1/T_2$ rates versus cumulative implantation doses. The inset shows the same plot in $T_2$ units. The $1/T_2$ is the rate with which electron spins of the NV$^{-}$ centers are homogeneously dephased, and it is proportional to nitrogen (P1 centers) concentration -- the main source of the spin dephasing~\cite{bauch_decoherence_2020}. The larger implantation dose, the larger population of the initially presented P1 centers can be converted into other kinds of defects~\cite{ashfold_nitrogen_2020}. Note that unlike as it is for the $T_1$ relaxation, the $T_2$ of the sample with the largest implantation dose is increased after the second annealing, which may be a result of P1 conversion into H3 centers~\cite{sumikura_highly_2020, deak_formation_2014} or it may be a result of a drop in concentration of possibly present vacancy chains, as at $\approx$1100~$^{\circ}$C their concentration is greatly reduced, effectively reducing the concentration of vacancy related paramagnetic defects~\cite{yamamoto_extending_2013}. In both cases, from the perspective of $T_2$ time, the annealing at temperature 1100~$^{\circ}$C is favorable.
In our case the three samples had initially the same concentrations of nitrogen; by introducing vacancies and creating NV centers we effectively decrease the P1 concentration, and by this we increase the $T_2$ time. The sensitivity of pulse magnetometry methods is usually limited by the $T_2$ time~\cite{barry_sensitivity_2020}, which in turn may be limited by the $T_1$ time, as $T_2$$\approx$$T_1$/2~\cite{bar-gill_solid-state_2013}. The observed drop in the $T_1$ time in practice does not affect the potential magnetic sensitivity because for our samples $T_1$ times is by three orders of magnitude larger than $T_2$ times.

\subsection{Estimates of concentrations}
We go further by using previously published dependency of $T_2$ on concentration of P1 centers~\cite{bauch_decoherence_2020} to indirectly estimate the concentration of P1 centers. We use an equation $1/T_2=x/T_{\rm NV-P1}$, where $T_{\rm NV-P1}$ is the P1-dominated NV decoherence time per unit concentration $x$. From the fit of the numerical simulation data we extract $T_{\rm NV-P1}$ = 80 $\upmu$s ppm. Note that we do not use the experimental dependence from the same Ref.~\cite{bauch_decoherence_2020} since it leads to concentrations of P1 centers that are much larger than a known initial concentration of  nitrogen, which is given by the manufacturer as $\approx$ 100 ppm. Concentrations of P1 centers estimated from the measured $T_2$  relaxation times are depicted on FIG.~\ref{Relaxations} c).
These estimates are used further to determine the total concentration of NV~$\approx$~NV$^-$ + NV$^0$ by subtracting the P1 concentration obtained from $T_2$ from the known initial nitrogen concentration, see FIG.~\ref{Relaxations} d). Then, by using a conservative value of the NV$^0$ to NV$^-$ charge-state conversion efficiency of 25~\% we estimate the NV$^-$ concentration, see FIG.~\ref{Relaxations} e). Based on the contrast measurement (FIG.~\ref{Fluorescence} b)) we assume that the conversion efficiency is the same for all three samples. We also assume that P1 centers may be converted into a tiny but not negligible concentration of nitrogen-containing defects (denoted as $\delta$) other than NV$^0$ and NV$^-$ centers.   

Similarly, we use  a measured dependency of $T_1$ on concentration of NV$^-$ centers, which is published in Ref.~\cite{jarmola_longitudinal_2015}.  
From the fit of the experimental data we found a linear equation $1/T_1=1/T_{\rm 1,other}+x/T_{\rm NV-NV}$, where $T_{\rm NV-NV}=0.08$~ms$\cdot$ppm is the dipole-dipole interactions driving relaxation time per unit concentration $x$ and the relaxation time $T_{\rm 1,other}=4.45$~ms accounts for other decoherence mechanisms. Concentrations of NV$^-$ centers estimated from the measured $T_1$  relaxation times are depicted on FIG.~\ref{Relaxations} f). This estimate leads to values about 10 ppm that is very similar to values estimated from $T_2$ relaxation, supporting our assumption about the charge-state conversion efficiency of 25~\%. 

However, slopes of the dependencies on the implantation dose are different. After the third annealing step, the estimates of NV$^-$ concentration derived from the $T_2$ have an increment by 4.3~ppm between the minimum and the maximum implantation doses, but the estimates derived from the $T_1$ have an increment only by 2.7~ppm between the same doses, see FIG.~\ref{Relaxations} e) and f)). The difference of the increments (slopes) has a physical meaning, and this could indicate one of the two (or combination of): a small concentration of N-containing defects $\delta$, which are neither P1 neither NV centers or a decrease in paramagnetic defects that are not related to nitrogen. Following the first hypothesis, the slopes after the first annealing at 800~$^{\circ}$C show a zero difference $\delta$ within error-bars.
That points to the H3 centers~\cite{sumikura_highly_2020, deak_formation_2014}, which formation is intensified at larger annealing temperatures.
The previously reported~\cite{sumikura_highly_2020,deak_formation_2014} concentration of H3 defects after an annealing at 1150~$^{\circ}$C is $\approx 1$\% of the NV center concentration, that is 0.6~ppm for 55~ppm of NV centers. Therefore, this expected concentration of H3 defects is of the same magnitude as the concentration $\delta=1.5\pm0.7$~ppm derived from difference of the estimates on FIG.~\ref{Relaxations} e) and f).

Following the second hypothesis, of a decrease of paramagnetic defects that are not related to nitrogen, one can argue that by increasing the annealing temperature to $\approx$1100~$^{\circ}$C the concentration of vacancy chains drops dramatically, effectively reducing the concentration of vacancy related paramagnetic defects~\cite{yamamoto_extending_2013}. This would enhance the $T_2$ relaxation time, but it would not change the $T_1$ relaxation time, as the $T_1$ time is sensitive only to the changes NV$^-$ bath. As a result the NV$^-$ concentration estimations (FIG.~\ref{Relaxations} e) and f)) from the $T_1$ and $T_2$ could be shifted because of the defects not related to nitrogen.

\begin{figure}
             \includegraphics[width=0.45\textwidth,valign=t]{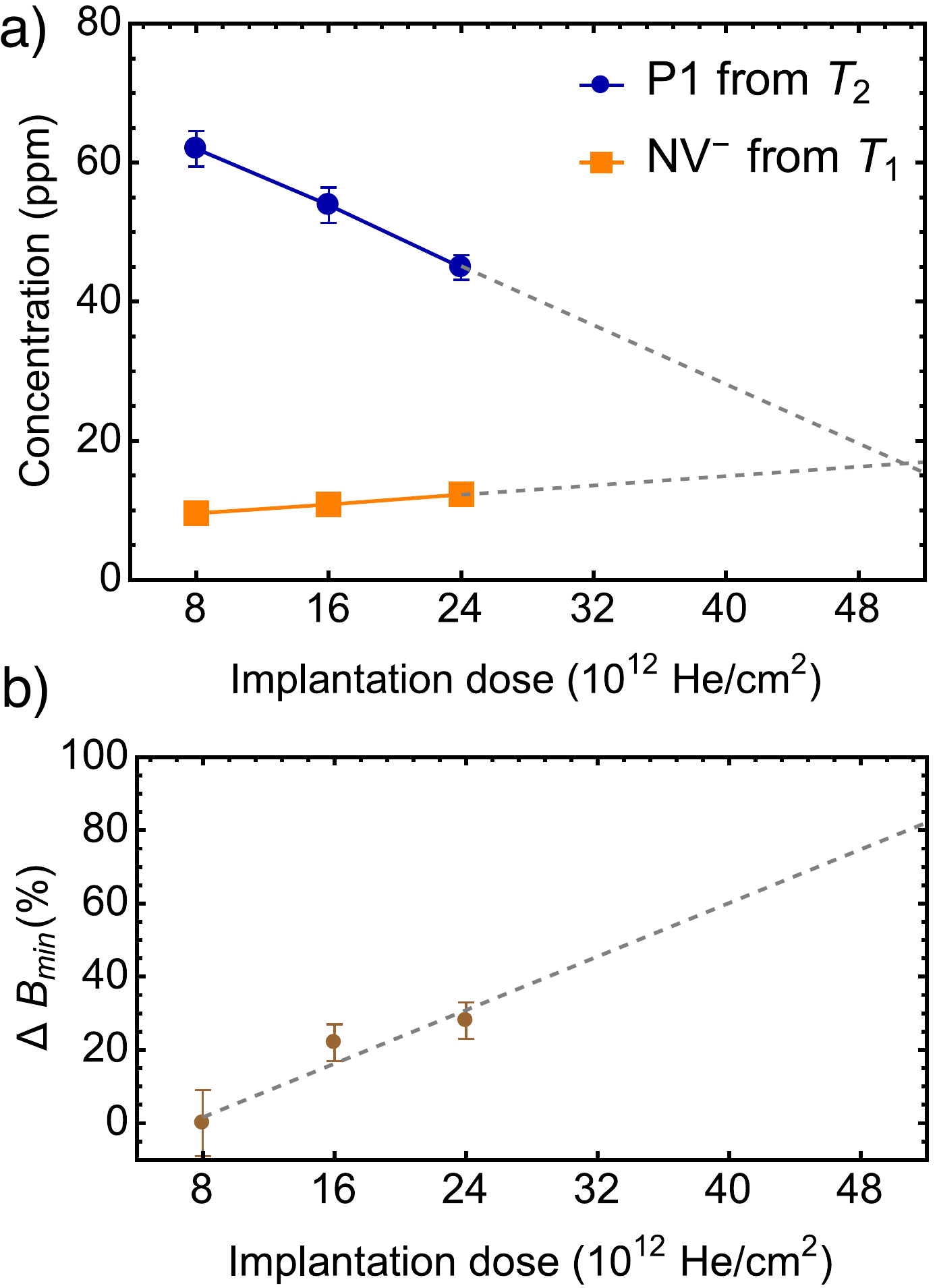}
             \caption{\textbf{Projected values: a)} 
             Linear extrapolations of P1 and NV$^-$ concentrations to higher implantation doses. The optimal dose is expected at $0.5\times10^{14}$~He$^+$/cm$^2$ where 
             the P1 concentration is equal to the NV$^-$ concentration. \textbf{b)} The linear extrapolation to higher implantation doses of relative improvement of sensitivity (minimum detectable magnetic field $B_{\rm min}$, see Eq.~\ref{eq:sen}).
        }
  \label{outlook}
\end{figure}

\section{Summary and outlook}
With this research we set out to find an optimal implantation dose and annealing parameters to maximize the sensitivity of an NV based sensor. Our efforts was focused on relatively cheap HPHT diamonds with high initial nitrogen concentration (100~ppm), as these kind of diamond based sensors would be of interest for mass-production of high sensitivity sensors. Our measurements clearly show that the strife for a higher sensitivity sensor (higher NV$^-$ concentration) not necessarily leads to the degradation of the sensor properties.

Since our data show a linear increase in NV$^-$ concentration upon increasing the He$^+$ implantation dose, we can conclude that from the sensitivity perspective it is lucrative to use thrice the ion implantation doses than reported in previously.
Assuming that at least one P1  center is needed as an electron donor for each NV$^-$, we estimate the maximum cumulative dose that could be used to saturate the NV$^-$ concentration for the $\approx$~200~nm thick layer with initial nitrogen concentration of 100~ppm. For this we fit our data with linear functions and extrapolate to a dose, where the P1 center concentration is equal to the NV$^-$ center concentration. The maximum cumulative dose obtained in this way is $\approx0.5~\times$~10$^{14}$~He$^+$/cm$^2$ (see Fig.~\ref{outlook} a)). This is also consistent with the estimations of He$^+$ dose and NV$^-$ concentrations for similar samples in Ref.~\cite{huang_diamond_2013}, which reports a sign of saturation at such a dose.

Similarly, we estimate a relative improvement of sensitivity for the dose of $0.5~\times~10^{14}$~He$^+$/cm$^2$ by interpolating the values of relative obtained from Eq.~\ref{eq:sen}, see Fig.~\ref{outlook} b). 
If we optimistically assume a linear growth of the sensitivity then we could expect a significant potential improvement up to 70\%. A half of this improvement is already achieved in this study. 

Dependencies of different measured characteristics on annealing suggest that the annealing only at 800~$^{\circ}$C does not deliver the optimal charge-state conversion efficiency, and only after the additional 2~h annealing at 1100~$^{\circ}$C the fluorescence reaches its maximum, and relaxation time $T_2$ reaches its extreme value. This might be connected to the reduction of the vacancy chain related paramagnetic defects observed at temperatures above 1100~$^{\circ}$C~\cite{yamamoto_extending_2013}, or it might be connected to conversion of P1 centres into H3 centers~\cite{sumikura_highly_2020} that also increases the $T_2$ time. Our results show that the average relative improvement of sensitivity between the first annealing at 800~$^{\circ}$C and the second annealing at 1100~$^{\circ}$C is 6.6~$\pm$~2.7~\%. 
While we do not see perspectives for further improvement of the sensitivity by adjusting the annealing procedure, we conclude that the annealing at 1100~$^{\circ}$C should not be neglected during fabrication of NV sensors. 

\section{Acknowledgements}

A. Berzins acknowledges support from Latvian Council of Science project lzp-2021/1-0379 "A novel solution for high magnetic field and high electric current stabilization using color centers in diamond", and LLC "MikroTik" donation project, administered by the UoL foundation, "Improvement of Magnetic field imaging system" for opportunity to significantly improve experimental setup as well as "Simulations for stimulation of science" for opportunity to acquire COMSOL licence. I. Fescenko acknowledges support from ERAF project 1.1.1.5/20/A/001 and I.F. and A.B. acknowledges support from LLC "MikroTik" donation project "Annealing furnace for the development of new nanometer-sized sensors and devices", administered by the University of Latvia Foundation.

\bibliography{references.bib}

\begin{thebibliography}{52}
\expandafter\ifx\csname natexlab\endcsname\relax\def\natexlab#1{#1}\fi
\expandafter\ifx\csname bibnamefont\endcsname\relax
  \def\bibnamefont#1{#1}\fi
\expandafter\ifx\csname bibfnamefont\endcsname\relax
  \def\bibfnamefont#1{#1}\fi
\expandafter\ifx\csname citenamefont\endcsname\relax
  \def\citenamefont#1{#1}\fi
\expandafter\ifx\csname url\endcsname\relax
  \def\url#1{\texttt{#1}}\fi
\expandafter\ifx\csname urlprefix\endcsname\relax\def\urlprefix{URL }\fi
\providecommand{\bibinfo}[2]{#2}
\providecommand{\eprint}[2][]{\url{#2}}

\bibitem[{\citenamefont{Ashfold et~al.}(2020)\citenamefont{Ashfold, Goss,
  Green, May, Newton, and Peaker}}]{ashfold_nitrogen_2020}
\bibinfo{author}{\bibfnamefont{M.~N.~R.} \bibnamefont{Ashfold}},
  \bibinfo{author}{\bibfnamefont{J.~P.} \bibnamefont{Goss}},
  \bibinfo{author}{\bibfnamefont{B.~L.} \bibnamefont{Green}},
  \bibinfo{author}{\bibfnamefont{P.~W.} \bibnamefont{May}},
  \bibinfo{author}{\bibfnamefont{M.~E.} \bibnamefont{Newton}},
  \bibnamefont{and} \bibinfo{author}{\bibfnamefont{C.~V.}
  \bibnamefont{Peaker}}, \bibinfo{journal}{Chemical Reviews}
  \textbf{\bibinfo{volume}{120}}, \bibinfo{pages}{5745} (\bibinfo{year}{2020}),
  ISSN \bibinfo{issn}{0009-2665, 1520-6890},
  \urlprefix\url{https://pubs.acs.org/doi/10.1021/acs.chemrev.9b00518}.

\bibitem[{\citenamefont{Barry et~al.}(2020)\citenamefont{Barry, Schloss, Bauch,
  Turner, Hart, Pham, and Walsworth}}]{barry_sensitivity_2020}
\bibinfo{author}{\bibfnamefont{J.~F.} \bibnamefont{Barry}},
  \bibinfo{author}{\bibfnamefont{J.~M.} \bibnamefont{Schloss}},
  \bibinfo{author}{\bibfnamefont{E.}~\bibnamefont{Bauch}},
  \bibinfo{author}{\bibfnamefont{M.~J.} \bibnamefont{Turner}},
  \bibinfo{author}{\bibfnamefont{C.~A.} \bibnamefont{Hart}},
  \bibinfo{author}{\bibfnamefont{L.~M.} \bibnamefont{Pham}}, \bibnamefont{and}
  \bibinfo{author}{\bibfnamefont{R.~L.} \bibnamefont{Walsworth}},
  \bibinfo{journal}{Reviews of Modern Physics} \textbf{\bibinfo{volume}{92}},
  \bibinfo{pages}{015004} (\bibinfo{year}{2020}), ISSN
  \bibinfo{issn}{0034-6861, 1539-0756},
  \urlprefix\url{https://link.aps.org/doi/10.1103/RevModPhys.92.015004}.

\bibitem[{\citenamefont{Wu et~al.}(2016)\citenamefont{Wu, Jelezko, Plenio, and
  Weil}}]{wu_diamond_2016}
\bibinfo{author}{\bibfnamefont{Y.}~\bibnamefont{Wu}},
  \bibinfo{author}{\bibfnamefont{F.}~\bibnamefont{Jelezko}},
  \bibinfo{author}{\bibfnamefont{M.~B.} \bibnamefont{Plenio}},
  \bibnamefont{and} \bibinfo{author}{\bibfnamefont{T.}~\bibnamefont{Weil}},
  \bibinfo{journal}{Angewandte Chemie International Edition}
  \textbf{\bibinfo{volume}{55}}, \bibinfo{pages}{6586} (\bibinfo{year}{2016}),
  ISSN \bibinfo{issn}{14337851},
  \urlprefix\url{http://doi.wiley.com/10.1002/anie.201506556}.

\bibitem[{\citenamefont{Chipaux et~al.}(2018)\citenamefont{Chipaux, van~der
  Laan, Hemelaar, Hasani, Zheng, and Schirhagl}}]{chipaux_nanodiamonds_2018}
\bibinfo{author}{\bibfnamefont{M.}~\bibnamefont{Chipaux}},
  \bibinfo{author}{\bibfnamefont{K.~J.} \bibnamefont{van~der Laan}},
  \bibinfo{author}{\bibfnamefont{S.~R.} \bibnamefont{Hemelaar}},
  \bibinfo{author}{\bibfnamefont{M.}~\bibnamefont{Hasani}},
  \bibinfo{author}{\bibfnamefont{T.}~\bibnamefont{Zheng}}, \bibnamefont{and}
  \bibinfo{author}{\bibfnamefont{R.}~\bibnamefont{Schirhagl}},
  \bibinfo{journal}{Small} \textbf{\bibinfo{volume}{14}},
  \bibinfo{pages}{1704263} (\bibinfo{year}{2018}), ISSN
  \bibinfo{issn}{16136810},
  \urlprefix\url{http://doi.wiley.com/10.1002/smll.201704263}.

\bibitem[{\citenamefont{Norman et~al.}(2020)\citenamefont{Norman, Majety, Wang,
  Casey, Curro, and Radulaski}}]{norman_novel_2020}
\bibinfo{author}{\bibfnamefont{V.~A.} \bibnamefont{Norman}},
  \bibinfo{author}{\bibfnamefont{S.}~\bibnamefont{Majety}},
  \bibinfo{author}{\bibfnamefont{Z.}~\bibnamefont{Wang}},
  \bibinfo{author}{\bibfnamefont{W.~H.} \bibnamefont{Casey}},
  \bibinfo{author}{\bibfnamefont{N.}~\bibnamefont{Curro}}, \bibnamefont{and}
  \bibinfo{author}{\bibfnamefont{M.}~\bibnamefont{Radulaski}},
  \bibinfo{journal}{InfoMat} p. \bibinfo{pages}{inf2.12128}
  (\bibinfo{year}{2020}), ISSN \bibinfo{issn}{2567-3165, 2567-3165},
  \urlprefix\url{https://onlinelibrary.wiley.com/doi/abs/10.1002/inf2.12128}.

\bibitem[{\citenamefont{Fu et~al.}(2020)\citenamefont{Fu, Iwata, Wickenbrock,
  and Budker}}]{fu_sensitive_2020}
\bibinfo{author}{\bibfnamefont{K.-M.~C.} \bibnamefont{Fu}},
  \bibinfo{author}{\bibfnamefont{G.~Z.} \bibnamefont{Iwata}},
  \bibinfo{author}{\bibfnamefont{A.}~\bibnamefont{Wickenbrock}},
  \bibnamefont{and} \bibinfo{author}{\bibfnamefont{D.}~\bibnamefont{Budker}},
  \bibinfo{journal}{AVS Quantum Science} \textbf{\bibinfo{volume}{2}},
  \bibinfo{pages}{044702} (\bibinfo{year}{2020}), ISSN
  \bibinfo{issn}{2639-0213},
  \urlprefix\url{http://avs.scitation.org/doi/10.1116/5.0025186}.

\bibitem[{\citenamefont{Abe and Sasaki}(2018)}]{abe_tutorial_2018}
\bibinfo{author}{\bibfnamefont{E.}~\bibnamefont{Abe}} \bibnamefont{and}
  \bibinfo{author}{\bibfnamefont{K.}~\bibnamefont{Sasaki}},
  \bibinfo{journal}{Journal of Applied Physics} \textbf{\bibinfo{volume}{123}},
  \bibinfo{pages}{161101} (\bibinfo{year}{2018}), ISSN
  \bibinfo{issn}{0021-8979, 1089-7550},
  \urlprefix\url{http://aip.scitation.org/doi/10.1063/1.5011231}.

\bibitem[{\citenamefont{Levine et~al.}(2019)\citenamefont{Levine, Turner,
  Kehayias, Hart, Langellier, Trubko, Glenn, Fu, and
  Walsworth}}]{levine_principles_2019}
\bibinfo{author}{\bibfnamefont{E.~V.} \bibnamefont{Levine}},
  \bibinfo{author}{\bibfnamefont{M.~J.} \bibnamefont{Turner}},
  \bibinfo{author}{\bibfnamefont{P.}~\bibnamefont{Kehayias}},
  \bibinfo{author}{\bibfnamefont{C.~A.} \bibnamefont{Hart}},
  \bibinfo{author}{\bibfnamefont{N.}~\bibnamefont{Langellier}},
  \bibinfo{author}{\bibfnamefont{R.}~\bibnamefont{Trubko}},
  \bibinfo{author}{\bibfnamefont{D.~R.} \bibnamefont{Glenn}},
  \bibinfo{author}{\bibfnamefont{R.~R.} \bibnamefont{Fu}}, \bibnamefont{and}
  \bibinfo{author}{\bibfnamefont{R.~L.} \bibnamefont{Walsworth}},
  \bibinfo{journal}{Nanophotonics} \textbf{\bibinfo{volume}{8}},
  \bibinfo{pages}{1945} (\bibinfo{year}{2019}), ISSN \bibinfo{issn}{2192-8614},
  \urlprefix\url{https://www.degruyter.com/document/doi/10.1515/nanoph-2019-0209/html}.

\bibitem[{\citenamefont{Rondin et~al.}(2014)\citenamefont{Rondin, Tetienne,
  Hingant, Roch, Maletinsky, and Jacques}}]{rondin_magnetometry_2014}
\bibinfo{author}{\bibfnamefont{L.}~\bibnamefont{Rondin}},
  \bibinfo{author}{\bibfnamefont{J.-P.} \bibnamefont{Tetienne}},
  \bibinfo{author}{\bibfnamefont{T.}~\bibnamefont{Hingant}},
  \bibinfo{author}{\bibfnamefont{J.-F.} \bibnamefont{Roch}},
  \bibinfo{author}{\bibfnamefont{P.}~\bibnamefont{Maletinsky}},
  \bibnamefont{and} \bibinfo{author}{\bibfnamefont{V.}~\bibnamefont{Jacques}},
  \bibinfo{journal}{Reports on Progress in Physics}
  \textbf{\bibinfo{volume}{77}}, \bibinfo{pages}{056503}
  (\bibinfo{year}{2014}), ISSN \bibinfo{issn}{0034-4885, 1361-6633},
  \urlprefix\url{https://iopscience.iop.org/article/10.1088/0034-4885/77/5/056503}.

\bibitem[{\citenamefont{Kehayias et~al.}(2017)\citenamefont{Kehayias, Jarmola,
  Mosavian, Fescenko, Benito, Laraoui, Smits, Bougas, Budker, Neumann
  et~al.}}]{kehayias_solution_2017}
\bibinfo{author}{\bibfnamefont{P.}~\bibnamefont{Kehayias}},
  \bibinfo{author}{\bibfnamefont{A.}~\bibnamefont{Jarmola}},
  \bibinfo{author}{\bibfnamefont{N.}~\bibnamefont{Mosavian}},
  \bibinfo{author}{\bibfnamefont{I.}~\bibnamefont{Fescenko}},
  \bibinfo{author}{\bibfnamefont{F.~M.} \bibnamefont{Benito}},
  \bibinfo{author}{\bibfnamefont{A.}~\bibnamefont{Laraoui}},
  \bibinfo{author}{\bibfnamefont{J.}~\bibnamefont{Smits}},
  \bibinfo{author}{\bibfnamefont{L.}~\bibnamefont{Bougas}},
  \bibinfo{author}{\bibfnamefont{D.}~\bibnamefont{Budker}},
  \bibinfo{author}{\bibfnamefont{A.}~\bibnamefont{Neumann}},
  \bibnamefont{et~al.}, \bibinfo{journal}{Nature Communications}
  \textbf{\bibinfo{volume}{8}}, \bibinfo{pages}{1} (\bibinfo{year}{2017}), ISSN
  \bibinfo{issn}{2041-1723}, \bibinfo{note}{number: 1 Publisher: Nature
  Publishing Group},
  \urlprefix\url{https://www.nature.com/articles/s41467-017-00266-4}.

\bibitem[{\citenamefont{Yamamoto et~al.}(2013)\citenamefont{Yamamoto, Umeda,
  Watanabe, Onoda, Markham, Twitchen, Naydenov, McGuinness, Teraji, Koizumi
  et~al.}}]{yamamoto_extending_2013}
\bibinfo{author}{\bibfnamefont{T.}~\bibnamefont{Yamamoto}},
  \bibinfo{author}{\bibfnamefont{T.}~\bibnamefont{Umeda}},
  \bibinfo{author}{\bibfnamefont{K.}~\bibnamefont{Watanabe}},
  \bibinfo{author}{\bibfnamefont{S.}~\bibnamefont{Onoda}},
  \bibinfo{author}{\bibfnamefont{M.~L.} \bibnamefont{Markham}},
  \bibinfo{author}{\bibfnamefont{D.~J.} \bibnamefont{Twitchen}},
  \bibinfo{author}{\bibfnamefont{B.}~\bibnamefont{Naydenov}},
  \bibinfo{author}{\bibfnamefont{L.~P.} \bibnamefont{McGuinness}},
  \bibinfo{author}{\bibfnamefont{T.}~\bibnamefont{Teraji}},
  \bibinfo{author}{\bibfnamefont{S.}~\bibnamefont{Koizumi}},
  \bibnamefont{et~al.}, \bibinfo{journal}{Physical Review B}
  \textbf{\bibinfo{volume}{88}}, \bibinfo{pages}{075206}
  (\bibinfo{year}{2013}), ISSN \bibinfo{issn}{1098-0121, 1550-235X},
  \urlprefix\url{https://link.aps.org/doi/10.1103/PhysRevB.88.075206}.

\bibitem[{\citenamefont{Naydenov
  et~al.}(2010{\natexlab{a}})\citenamefont{Naydenov, Richter, Beck, Steiner,
  Neumann, Balasubramanian, Achard, Jelezko, Wrachtrup, and
  Kalish}}]{naydenov_enhanced_2010}
\bibinfo{author}{\bibfnamefont{B.}~\bibnamefont{Naydenov}},
  \bibinfo{author}{\bibfnamefont{V.}~\bibnamefont{Richter}},
  \bibinfo{author}{\bibfnamefont{J.}~\bibnamefont{Beck}},
  \bibinfo{author}{\bibfnamefont{M.}~\bibnamefont{Steiner}},
  \bibinfo{author}{\bibfnamefont{P.}~\bibnamefont{Neumann}},
  \bibinfo{author}{\bibfnamefont{G.}~\bibnamefont{Balasubramanian}},
  \bibinfo{author}{\bibfnamefont{J.}~\bibnamefont{Achard}},
  \bibinfo{author}{\bibfnamefont{F.}~\bibnamefont{Jelezko}},
  \bibinfo{author}{\bibfnamefont{J.}~\bibnamefont{Wrachtrup}},
  \bibnamefont{and} \bibinfo{author}{\bibfnamefont{R.}~\bibnamefont{Kalish}},
  \bibinfo{journal}{Applied Physics Letters} \textbf{\bibinfo{volume}{96}},
  \bibinfo{pages}{163108} (\bibinfo{year}{2010}{\natexlab{a}}), ISSN
  \bibinfo{issn}{0003-6951, 1077-3118},
  \urlprefix\url{http://aip.scitation.org/doi/10.1063/1.3409221}.

\bibitem[{\citenamefont{Rabeau et~al.}(2006)\citenamefont{Rabeau, Reichart,
  Tamanyan, Jamieson, Prawer, Jelezko, Gaebel, Popa, Domhan, and
  Wrachtrup}}]{rabeau_implantation_2006}
\bibinfo{author}{\bibfnamefont{J.~R.} \bibnamefont{Rabeau}},
  \bibinfo{author}{\bibfnamefont{P.}~\bibnamefont{Reichart}},
  \bibinfo{author}{\bibfnamefont{G.}~\bibnamefont{Tamanyan}},
  \bibinfo{author}{\bibfnamefont{D.~N.} \bibnamefont{Jamieson}},
  \bibinfo{author}{\bibfnamefont{S.}~\bibnamefont{Prawer}},
  \bibinfo{author}{\bibfnamefont{F.}~\bibnamefont{Jelezko}},
  \bibinfo{author}{\bibfnamefont{T.}~\bibnamefont{Gaebel}},
  \bibinfo{author}{\bibfnamefont{I.}~\bibnamefont{Popa}},
  \bibinfo{author}{\bibfnamefont{M.}~\bibnamefont{Domhan}}, \bibnamefont{and}
  \bibinfo{author}{\bibfnamefont{J.}~\bibnamefont{Wrachtrup}},
  \bibinfo{journal}{Applied Physics Letters} \textbf{\bibinfo{volume}{88}},
  \bibinfo{pages}{023113} (\bibinfo{year}{2006}), ISSN
  \bibinfo{issn}{0003-6951, 1077-3118},
  \urlprefix\url{http://aip.scitation.org/doi/10.1063/1.2158700}.

\bibitem[{\citenamefont{Meijer et~al.}(2005)\citenamefont{Meijer, Burchard,
  Domhan, Wittmann, Gaebel, Popa, Jelezko, and
  Wrachtrup}}]{meijer_generation_2005}
\bibinfo{author}{\bibfnamefont{J.}~\bibnamefont{Meijer}},
  \bibinfo{author}{\bibfnamefont{B.}~\bibnamefont{Burchard}},
  \bibinfo{author}{\bibfnamefont{M.}~\bibnamefont{Domhan}},
  \bibinfo{author}{\bibfnamefont{C.}~\bibnamefont{Wittmann}},
  \bibinfo{author}{\bibfnamefont{T.}~\bibnamefont{Gaebel}},
  \bibinfo{author}{\bibfnamefont{I.}~\bibnamefont{Popa}},
  \bibinfo{author}{\bibfnamefont{F.}~\bibnamefont{Jelezko}}, \bibnamefont{and}
  \bibinfo{author}{\bibfnamefont{J.}~\bibnamefont{Wrachtrup}},
  \bibinfo{journal}{Applied Physics Letters} \textbf{\bibinfo{volume}{87}},
  \bibinfo{pages}{261909} (\bibinfo{year}{2005}), ISSN
  \bibinfo{issn}{0003-6951, 1077-3118},
  \urlprefix\url{http://aip.scitation.org/doi/10.1063/1.2103389}.

\bibitem[{\citenamefont{Fu et~al.}(2010)\citenamefont{Fu, Santori, Barclay, and
  Beausoleil}}]{fu_conversion_2010}
\bibinfo{author}{\bibfnamefont{K.-M.~C.} \bibnamefont{Fu}},
  \bibinfo{author}{\bibfnamefont{C.}~\bibnamefont{Santori}},
  \bibinfo{author}{\bibfnamefont{P.~E.} \bibnamefont{Barclay}},
  \bibnamefont{and} \bibinfo{author}{\bibfnamefont{R.~G.}
  \bibnamefont{Beausoleil}}, \bibinfo{journal}{Applied Physics Letters}
  \textbf{\bibinfo{volume}{96}}, \bibinfo{pages}{121907}
  (\bibinfo{year}{2010}), ISSN \bibinfo{issn}{0003-6951, 1077-3118},
  \urlprefix\url{http://aip.scitation.org/doi/10.1063/1.3364135}.

\bibitem[{\citenamefont{Aharonovich et~al.}(2009)\citenamefont{Aharonovich,
  Santori, Fairchild, Orwa, Ganesan, Fu, Beausoleil, Greentree, and
  Prawer}}]{aharonovich_producing_2009}
\bibinfo{author}{\bibfnamefont{I.}~\bibnamefont{Aharonovich}},
  \bibinfo{author}{\bibfnamefont{C.}~\bibnamefont{Santori}},
  \bibinfo{author}{\bibfnamefont{B.~A.} \bibnamefont{Fairchild}},
  \bibinfo{author}{\bibfnamefont{J.}~\bibnamefont{Orwa}},
  \bibinfo{author}{\bibfnamefont{K.}~\bibnamefont{Ganesan}},
  \bibinfo{author}{\bibfnamefont{K.-M.~C.} \bibnamefont{Fu}},
  \bibinfo{author}{\bibfnamefont{R.~G.} \bibnamefont{Beausoleil}},
  \bibinfo{author}{\bibfnamefont{A.~D.} \bibnamefont{Greentree}},
  \bibnamefont{and} \bibinfo{author}{\bibfnamefont{S.}~\bibnamefont{Prawer}},
  \bibinfo{journal}{Journal of Applied Physics} \textbf{\bibinfo{volume}{106}},
  \bibinfo{pages}{124904} (\bibinfo{year}{2009}), ISSN
  \bibinfo{issn}{0021-8979, 1089-7550},
  \urlprefix\url{http://aip.scitation.org/doi/10.1063/1.3271579}.

\bibitem[{\citenamefont{Naydenov
  et~al.}(2010{\natexlab{b}})\citenamefont{Naydenov, Reinhard, Lämmle,
  Richter, Kalish, D’Haenens-Johansson, Newton, Jelezko, and
  Wrachtrup}}]{naydenov_increasing_2010}
\bibinfo{author}{\bibfnamefont{B.}~\bibnamefont{Naydenov}},
  \bibinfo{author}{\bibfnamefont{F.}~\bibnamefont{Reinhard}},
  \bibinfo{author}{\bibfnamefont{A.}~\bibnamefont{Lämmle}},
  \bibinfo{author}{\bibfnamefont{V.}~\bibnamefont{Richter}},
  \bibinfo{author}{\bibfnamefont{R.}~\bibnamefont{Kalish}},
  \bibinfo{author}{\bibfnamefont{U.~F.~S.}
  \bibnamefont{D’Haenens-Johansson}},
  \bibinfo{author}{\bibfnamefont{M.}~\bibnamefont{Newton}},
  \bibinfo{author}{\bibfnamefont{F.}~\bibnamefont{Jelezko}}, \bibnamefont{and}
  \bibinfo{author}{\bibfnamefont{J.}~\bibnamefont{Wrachtrup}},
  \bibinfo{journal}{Applied Physics Letters} \textbf{\bibinfo{volume}{97}},
  \bibinfo{pages}{242511} (\bibinfo{year}{2010}{\natexlab{b}}), ISSN
  \bibinfo{issn}{0003-6951, 1077-3118},
  \urlprefix\url{http://aip.scitation.org/doi/10.1063/1.3527975}.

\bibitem[{\citenamefont{Luo et~al.}(2022)\citenamefont{Luo, Lindner, Langer,
  Cimalla, Vidal, Hahl, Schreyvogel, Onoda, Ishii, Ohshima
  et~al.}}]{luo_creation_2022}
\bibinfo{author}{\bibfnamefont{T.}~\bibnamefont{Luo}},
  \bibinfo{author}{\bibfnamefont{L.}~\bibnamefont{Lindner}},
  \bibinfo{author}{\bibfnamefont{J.}~\bibnamefont{Langer}},
  \bibinfo{author}{\bibfnamefont{V.}~\bibnamefont{Cimalla}},
  \bibinfo{author}{\bibfnamefont{X.}~\bibnamefont{Vidal}},
  \bibinfo{author}{\bibfnamefont{F.}~\bibnamefont{Hahl}},
  \bibinfo{author}{\bibfnamefont{C.}~\bibnamefont{Schreyvogel}},
  \bibinfo{author}{\bibfnamefont{S.}~\bibnamefont{Onoda}},
  \bibinfo{author}{\bibfnamefont{S.}~\bibnamefont{Ishii}},
  \bibinfo{author}{\bibfnamefont{T.}~\bibnamefont{Ohshima}},
  \bibnamefont{et~al.}, \bibinfo{journal}{New Journal of Physics}
  (\bibinfo{year}{2022}), ISSN \bibinfo{issn}{1367-2630},
  \urlprefix\url{https://iopscience.iop.org/article/10.1088/1367-2630/ac58b6}.

\bibitem[{\citenamefont{Bassett et~al.}(2011)\citenamefont{Bassett, Heremans,
  Yale, Buckley, and Awschalom}}]{bassett_electrical_2011}
\bibinfo{author}{\bibfnamefont{L.~C.} \bibnamefont{Bassett}},
  \bibinfo{author}{\bibfnamefont{F.~J.} \bibnamefont{Heremans}},
  \bibinfo{author}{\bibfnamefont{C.~G.} \bibnamefont{Yale}},
  \bibinfo{author}{\bibfnamefont{B.~B.} \bibnamefont{Buckley}},
  \bibnamefont{and} \bibinfo{author}{\bibfnamefont{D.~D.}
  \bibnamefont{Awschalom}}, \bibinfo{journal}{Physical Review Letters}
  \textbf{\bibinfo{volume}{107}}, \bibinfo{pages}{266403}
  (\bibinfo{year}{2011}), ISSN \bibinfo{issn}{0031-9007, 1079-7114},
  \urlprefix\url{https://link.aps.org/doi/10.1103/PhysRevLett.107.266403}.

\bibitem[{\citenamefont{Mrózek et~al.}(2021)\citenamefont{Mrózek,
  Wojciechowski, and Gawlik}}]{mrozek_characterization_2021}
\bibinfo{author}{\bibfnamefont{M.}~\bibnamefont{Mrózek}},
  \bibinfo{author}{\bibfnamefont{A.~M.} \bibnamefont{Wojciechowski}},
  \bibnamefont{and} \bibinfo{author}{\bibfnamefont{W.}~\bibnamefont{Gawlik}},
  \bibinfo{journal}{Diamond and Related Materials}
  \textbf{\bibinfo{volume}{120}}, \bibinfo{pages}{108689}
  (\bibinfo{year}{2021}), ISSN \bibinfo{issn}{09259635},
  \urlprefix\url{https://linkinghub.elsevier.com/retrieve/pii/S0925963521004520}.

\bibitem[{\citenamefont{Bogdanov et~al.}(2021)\citenamefont{Bogdanov,
  Gorbachev, Radishev, Vikharev, Lobaev, Bolshedvorskii, Soshenko, Gusev,
  Tatarskiy, and Akimov}}]{bogdanov_investigation_2021}
\bibinfo{author}{\bibfnamefont{S.}~\bibnamefont{Bogdanov}},
  \bibinfo{author}{\bibfnamefont{A.}~\bibnamefont{Gorbachev}},
  \bibinfo{author}{\bibfnamefont{D.}~\bibnamefont{Radishev}},
  \bibinfo{author}{\bibfnamefont{A.}~\bibnamefont{Vikharev}},
  \bibinfo{author}{\bibfnamefont{M.}~\bibnamefont{Lobaev}},
  \bibinfo{author}{\bibfnamefont{S.}~\bibnamefont{Bolshedvorskii}},
  \bibinfo{author}{\bibfnamefont{V.}~\bibnamefont{Soshenko}},
  \bibinfo{author}{\bibfnamefont{S.}~\bibnamefont{Gusev}},
  \bibinfo{author}{\bibfnamefont{D.}~\bibnamefont{Tatarskiy}},
  \bibnamefont{and} \bibinfo{author}{\bibfnamefont{A.}~\bibnamefont{Akimov}},
  \bibinfo{journal}{physica status solidi (RRL) – Rapid Research Letters}
  \textbf{\bibinfo{volume}{15}}, \bibinfo{pages}{2000550}
  (\bibinfo{year}{2021}), ISSN \bibinfo{issn}{1862-6254, 1862-6270},
  \urlprefix\url{https://onlinelibrary.wiley.com/doi/10.1002/pssr.202000550}.

\bibitem[{\citenamefont{Jarmola et~al.}(2015)\citenamefont{Jarmola, Berzins,
  Smits, Smits, Prikulis, Gahbauer, Ferber, Erts, Auzinsh, and
  Budker}}]{jarmola_longitudinal_2015}
\bibinfo{author}{\bibfnamefont{A.}~\bibnamefont{Jarmola}},
  \bibinfo{author}{\bibfnamefont{A.}~\bibnamefont{Berzins}},
  \bibinfo{author}{\bibfnamefont{J.}~\bibnamefont{Smits}},
  \bibinfo{author}{\bibfnamefont{K.}~\bibnamefont{Smits}},
  \bibinfo{author}{\bibfnamefont{J.}~\bibnamefont{Prikulis}},
  \bibinfo{author}{\bibfnamefont{F.}~\bibnamefont{Gahbauer}},
  \bibinfo{author}{\bibfnamefont{R.}~\bibnamefont{Ferber}},
  \bibinfo{author}{\bibfnamefont{D.}~\bibnamefont{Erts}},
  \bibinfo{author}{\bibfnamefont{M.}~\bibnamefont{Auzinsh}}, \bibnamefont{and}
  \bibinfo{author}{\bibfnamefont{D.}~\bibnamefont{Budker}},
  \bibinfo{journal}{Applied Physics Letters} \textbf{\bibinfo{volume}{107}},
  \bibinfo{pages}{242403} (\bibinfo{year}{2015}), ISSN
  \bibinfo{issn}{0003-6951, 1077-3118},
  \urlprefix\url{http://aip.scitation.org/doi/10.1063/1.4937489}.

\bibitem[{\citenamefont{Kurita et~al.}(2021)\citenamefont{Kurita, Shimotsuma,
  Fujiwara, Fujie, Mizuochi, Shimizu, and Miura}}]{kurita_direct_2021}
\bibinfo{author}{\bibfnamefont{T.}~\bibnamefont{Kurita}},
  \bibinfo{author}{\bibfnamefont{Y.}~\bibnamefont{Shimotsuma}},
  \bibinfo{author}{\bibfnamefont{M.}~\bibnamefont{Fujiwara}},
  \bibinfo{author}{\bibfnamefont{M.}~\bibnamefont{Fujie}},
  \bibinfo{author}{\bibfnamefont{N.}~\bibnamefont{Mizuochi}},
  \bibinfo{author}{\bibfnamefont{M.}~\bibnamefont{Shimizu}}, \bibnamefont{and}
  \bibinfo{author}{\bibfnamefont{K.}~\bibnamefont{Miura}},
  \bibinfo{journal}{Applied Physics Letters} \textbf{\bibinfo{volume}{118}},
  \bibinfo{pages}{214001} (\bibinfo{year}{2021}), ISSN
  \bibinfo{issn}{0003-6951, 1077-3118},
  \urlprefix\url{https://aip.scitation.org/doi/10.1063/5.0049953}.

\bibitem[{\citenamefont{Chen et~al.}(2017)\citenamefont{Chen, Salter, Knauer,
  Weng, Frangeskou, Stephen, Ishmael, Dolan, Johnson, Green
  et~al.}}]{chen_laser_2017}
\bibinfo{author}{\bibfnamefont{Y.-C.} \bibnamefont{Chen}},
  \bibinfo{author}{\bibfnamefont{P.~S.} \bibnamefont{Salter}},
  \bibinfo{author}{\bibfnamefont{S.}~\bibnamefont{Knauer}},
  \bibinfo{author}{\bibfnamefont{L.}~\bibnamefont{Weng}},
  \bibinfo{author}{\bibfnamefont{A.~C.} \bibnamefont{Frangeskou}},
  \bibinfo{author}{\bibfnamefont{C.~J.} \bibnamefont{Stephen}},
  \bibinfo{author}{\bibfnamefont{S.~N.} \bibnamefont{Ishmael}},
  \bibinfo{author}{\bibfnamefont{P.~R.} \bibnamefont{Dolan}},
  \bibinfo{author}{\bibfnamefont{S.}~\bibnamefont{Johnson}},
  \bibinfo{author}{\bibfnamefont{B.~L.} \bibnamefont{Green}},
  \bibnamefont{et~al.}, \bibinfo{journal}{Nature Photonics}
  \textbf{\bibinfo{volume}{11}}, \bibinfo{pages}{77} (\bibinfo{year}{2017}),
  ISSN \bibinfo{issn}{1749-4885, 1749-4893},
  \urlprefix\url{http://www.nature.com/articles/nphoton.2016.234}.

\bibitem[{\citenamefont{Giri et~al.}(2018)\citenamefont{Giri, Gorrini,
  Dorigoni, Avalos, Cazzanelli, Tambalo, and Bifone}}]{giri_coupled_2018}
\bibinfo{author}{\bibfnamefont{R.}~\bibnamefont{Giri}},
  \bibinfo{author}{\bibfnamefont{F.}~\bibnamefont{Gorrini}},
  \bibinfo{author}{\bibfnamefont{C.}~\bibnamefont{Dorigoni}},
  \bibinfo{author}{\bibfnamefont{C.~E.} \bibnamefont{Avalos}},
  \bibinfo{author}{\bibfnamefont{M.}~\bibnamefont{Cazzanelli}},
  \bibinfo{author}{\bibfnamefont{S.}~\bibnamefont{Tambalo}}, \bibnamefont{and}
  \bibinfo{author}{\bibfnamefont{A.}~\bibnamefont{Bifone}},
  \bibinfo{journal}{Physical Review B} \textbf{\bibinfo{volume}{98}},
  \bibinfo{pages}{045401} (\bibinfo{year}{2018}), ISSN
  \bibinfo{issn}{2469-9950, 2469-9969},
  \urlprefix\url{https://link.aps.org/doi/10.1103/PhysRevB.98.045401}.

\bibitem[{\citenamefont{Wolf et~al.}(2015)\citenamefont{Wolf, Neumann,
  Nakamura, Sumiya, Ohshima, Isoya, and Wrachtrup}}]{wolf_subpicotesla_2015}
\bibinfo{author}{\bibfnamefont{T.}~\bibnamefont{Wolf}},
  \bibinfo{author}{\bibfnamefont{P.}~\bibnamefont{Neumann}},
  \bibinfo{author}{\bibfnamefont{K.}~\bibnamefont{Nakamura}},
  \bibinfo{author}{\bibfnamefont{H.}~\bibnamefont{Sumiya}},
  \bibinfo{author}{\bibfnamefont{T.}~\bibnamefont{Ohshima}},
  \bibinfo{author}{\bibfnamefont{J.}~\bibnamefont{Isoya}}, \bibnamefont{and}
  \bibinfo{author}{\bibfnamefont{J.}~\bibnamefont{Wrachtrup}},
  \bibinfo{journal}{Physical Review X} \textbf{\bibinfo{volume}{5}},
  \bibinfo{pages}{041001} (\bibinfo{year}{2015}), ISSN
  \bibinfo{issn}{2160-3308},
  \urlprefix\url{https://link.aps.org/doi/10.1103/PhysRevX.5.041001}.

\bibitem[{\citenamefont{Fescenko et~al.}(2019)\citenamefont{Fescenko, Laraoui,
  Smits, Mosavian, Kehayias, Seto, Bougas, Jarmola, and
  Acosta}}]{fescenko_diamond_2019}
\bibinfo{author}{\bibfnamefont{I.}~\bibnamefont{Fescenko}},
  \bibinfo{author}{\bibfnamefont{A.}~\bibnamefont{Laraoui}},
  \bibinfo{author}{\bibfnamefont{J.}~\bibnamefont{Smits}},
  \bibinfo{author}{\bibfnamefont{N.}~\bibnamefont{Mosavian}},
  \bibinfo{author}{\bibfnamefont{P.}~\bibnamefont{Kehayias}},
  \bibinfo{author}{\bibfnamefont{J.}~\bibnamefont{Seto}},
  \bibinfo{author}{\bibfnamefont{L.}~\bibnamefont{Bougas}},
  \bibinfo{author}{\bibfnamefont{A.}~\bibnamefont{Jarmola}}, \bibnamefont{and}
  \bibinfo{author}{\bibfnamefont{V.~M.} \bibnamefont{Acosta}},
  \bibinfo{journal}{Physical Review Applied} \textbf{\bibinfo{volume}{11}},
  \bibinfo{pages}{034029} (\bibinfo{year}{2019}), \bibinfo{note}{publisher:
  American Physical Society},
  \urlprefix\url{https://link.aps.org/doi/10.1103/PhysRevApplied.11.034029}.

\bibitem[{\citenamefont{Berzins
  et~al.}(2021{\natexlab{a}})\citenamefont{Berzins, Smits, and
  Petruhins}}]{berzins_characterization_2021}
\bibinfo{author}{\bibfnamefont{A.}~\bibnamefont{Berzins}},
  \bibinfo{author}{\bibfnamefont{J.}~\bibnamefont{Smits}}, \bibnamefont{and}
  \bibinfo{author}{\bibfnamefont{A.}~\bibnamefont{Petruhins}},
  \bibinfo{journal}{Materials Chemistry and Physics}
  \textbf{\bibinfo{volume}{267}}, \bibinfo{pages}{124617}
  (\bibinfo{year}{2021}{\natexlab{a}}), ISSN \bibinfo{issn}{02540584},
  \urlprefix\url{https://linkinghub.elsevier.com/retrieve/pii/S0254058421004004}.

\bibitem[{\citenamefont{Huang et~al.}(2013)\citenamefont{Huang, Li, Santori,
  Acosta, Faraon, Ishikawa, Wu, Winston, Williams, and
  Beausoleil}}]{huang_diamond_2013}
\bibinfo{author}{\bibfnamefont{Z.}~\bibnamefont{Huang}},
  \bibinfo{author}{\bibfnamefont{W.-D.} \bibnamefont{Li}},
  \bibinfo{author}{\bibfnamefont{C.}~\bibnamefont{Santori}},
  \bibinfo{author}{\bibfnamefont{V.~M.} \bibnamefont{Acosta}},
  \bibinfo{author}{\bibfnamefont{A.}~\bibnamefont{Faraon}},
  \bibinfo{author}{\bibfnamefont{T.}~\bibnamefont{Ishikawa}},
  \bibinfo{author}{\bibfnamefont{W.}~\bibnamefont{Wu}},
  \bibinfo{author}{\bibfnamefont{D.}~\bibnamefont{Winston}},
  \bibinfo{author}{\bibfnamefont{R.~S.} \bibnamefont{Williams}},
  \bibnamefont{and} \bibinfo{author}{\bibfnamefont{R.~G.}
  \bibnamefont{Beausoleil}}, \bibinfo{journal}{Applied Physics Letters}
  \textbf{\bibinfo{volume}{103}}, \bibinfo{pages}{081906}
  (\bibinfo{year}{2013}), ISSN \bibinfo{issn}{0003-6951, 1077-3118},
  \urlprefix\url{http://aip.scitation.org/doi/10.1063/1.4819339}.

\bibitem[{\citenamefont{Kleinsasser et~al.}(2016)\citenamefont{Kleinsasser,
  Stanfield, Banks, Zhu, Li, Acosta, Watanabe, Itoh, and
  Fu}}]{kleinsasser_high_2016}
\bibinfo{author}{\bibfnamefont{E.~E.} \bibnamefont{Kleinsasser}},
  \bibinfo{author}{\bibfnamefont{M.~M.} \bibnamefont{Stanfield}},
  \bibinfo{author}{\bibfnamefont{J.~K.~Q.} \bibnamefont{Banks}},
  \bibinfo{author}{\bibfnamefont{Z.}~\bibnamefont{Zhu}},
  \bibinfo{author}{\bibfnamefont{W.-D.} \bibnamefont{Li}},
  \bibinfo{author}{\bibfnamefont{V.~M.} \bibnamefont{Acosta}},
  \bibinfo{author}{\bibfnamefont{H.}~\bibnamefont{Watanabe}},
  \bibinfo{author}{\bibfnamefont{K.~M.} \bibnamefont{Itoh}}, \bibnamefont{and}
  \bibinfo{author}{\bibfnamefont{K.-M.~C.} \bibnamefont{Fu}},
  \bibinfo{journal}{Applied Physics Letters} \textbf{\bibinfo{volume}{108}},
  \bibinfo{pages}{202401} (\bibinfo{year}{2016}), ISSN
  \bibinfo{issn}{0003-6951, 1077-3118},
  \urlprefix\url{http://aip.scitation.org/doi/10.1063/1.4949357}.

\bibitem[{\citenamefont{Fávaro~de Oliveira
  et~al.}(2016)\citenamefont{Fávaro~de Oliveira, Momenzadeh, Antonov, Scharpf,
  Osterkamp, Naydenov, Jelezko, Denisenko, and
  Wrachtrup}}]{favaro_de_oliveira_toward_2016}
\bibinfo{author}{\bibfnamefont{F.}~\bibnamefont{Fávaro~de Oliveira}},
  \bibinfo{author}{\bibfnamefont{S.~A.} \bibnamefont{Momenzadeh}},
  \bibinfo{author}{\bibfnamefont{D.}~\bibnamefont{Antonov}},
  \bibinfo{author}{\bibfnamefont{J.}~\bibnamefont{Scharpf}},
  \bibinfo{author}{\bibfnamefont{C.}~\bibnamefont{Osterkamp}},
  \bibinfo{author}{\bibfnamefont{B.}~\bibnamefont{Naydenov}},
  \bibinfo{author}{\bibfnamefont{F.}~\bibnamefont{Jelezko}},
  \bibinfo{author}{\bibfnamefont{A.}~\bibnamefont{Denisenko}},
  \bibnamefont{and}
  \bibinfo{author}{\bibfnamefont{J.}~\bibnamefont{Wrachtrup}},
  \bibinfo{journal}{Nano Letters} \textbf{\bibinfo{volume}{16}},
  \bibinfo{pages}{2228} (\bibinfo{year}{2016}), ISSN \bibinfo{issn}{1530-6984,
  1530-6992},
  \urlprefix\url{https://pubs.acs.org/doi/10.1021/acs.nanolett.5b04511}.

\bibitem[{\citenamefont{Berzins
  et~al.}(2021{\natexlab{b}})\citenamefont{Berzins, Smits, Petruhins, and
  Grube}}]{berzins_surface_2021}
\bibinfo{author}{\bibfnamefont{A.}~\bibnamefont{Berzins}},
  \bibinfo{author}{\bibfnamefont{J.}~\bibnamefont{Smits}},
  \bibinfo{author}{\bibfnamefont{A.}~\bibnamefont{Petruhins}},
  \bibnamefont{and} \bibinfo{author}{\bibfnamefont{H.}~\bibnamefont{Grube}},
  \bibinfo{journal}{Materials Chemistry and Physics}
  \textbf{\bibinfo{volume}{272}}, \bibinfo{pages}{124972}
  (\bibinfo{year}{2021}{\natexlab{b}}), ISSN \bibinfo{issn}{02540584},
  \urlprefix\url{https://linkinghub.elsevier.com/retrieve/pii/S0254058421007550}.

\bibitem[{\citenamefont{Sumikura et~al.}(2020)\citenamefont{Sumikura, Hirama,
  Nishiguchi, Shinya, and Notomi}}]{sumikura_highly_2020}
\bibinfo{author}{\bibfnamefont{H.}~\bibnamefont{Sumikura}},
  \bibinfo{author}{\bibfnamefont{K.}~\bibnamefont{Hirama}},
  \bibinfo{author}{\bibfnamefont{K.}~\bibnamefont{Nishiguchi}},
  \bibinfo{author}{\bibfnamefont{A.}~\bibnamefont{Shinya}}, \bibnamefont{and}
  \bibinfo{author}{\bibfnamefont{M.}~\bibnamefont{Notomi}},
  \bibinfo{journal}{APL Materials} \textbf{\bibinfo{volume}{8}},
  \bibinfo{pages}{031113} (\bibinfo{year}{2020}), ISSN
  \bibinfo{issn}{2166-532X},
  \urlprefix\url{http://aip.scitation.org/doi/10.1063/5.0001922}.

\bibitem[{\citenamefont{Havlik et~al.}(2013)\citenamefont{Havlik, Petrakova,
  Rehor, Petrak, Gulka, Stursa, Kucka, Ralis, Rendler, Lee
  et~al.}}]{havlik_boosting_2013}
\bibinfo{author}{\bibfnamefont{J.}~\bibnamefont{Havlik}},
  \bibinfo{author}{\bibfnamefont{V.}~\bibnamefont{Petrakova}},
  \bibinfo{author}{\bibfnamefont{I.}~\bibnamefont{Rehor}},
  \bibinfo{author}{\bibfnamefont{V.}~\bibnamefont{Petrak}},
  \bibinfo{author}{\bibfnamefont{M.}~\bibnamefont{Gulka}},
  \bibinfo{author}{\bibfnamefont{J.}~\bibnamefont{Stursa}},
  \bibinfo{author}{\bibfnamefont{J.}~\bibnamefont{Kucka}},
  \bibinfo{author}{\bibfnamefont{J.}~\bibnamefont{Ralis}},
  \bibinfo{author}{\bibfnamefont{T.}~\bibnamefont{Rendler}},
  \bibinfo{author}{\bibfnamefont{S.-Y.} \bibnamefont{Lee}},
  \bibnamefont{et~al.}, \bibinfo{journal}{Nanoscale}
  \textbf{\bibinfo{volume}{5}}, \bibinfo{pages}{3208} (\bibinfo{year}{2013}),
  ISSN \bibinfo{issn}{2040-3364, 2040-3372},
  \urlprefix\url{http://xlink.rsc.org/?DOI=c2nr32778c}.

\bibitem[{\citenamefont{McCloskey et~al.}(2014)\citenamefont{McCloskey, Fox,
  O'Hara, Usov, Scanlan, McEvoy, Duesberg, Cross, Zhang, and
  Donegan}}]{mccloskey_helium_2014}
\bibinfo{author}{\bibfnamefont{D.}~\bibnamefont{McCloskey}},
  \bibinfo{author}{\bibfnamefont{D.}~\bibnamefont{Fox}},
  \bibinfo{author}{\bibfnamefont{N.}~\bibnamefont{O'Hara}},
  \bibinfo{author}{\bibfnamefont{V.}~\bibnamefont{Usov}},
  \bibinfo{author}{\bibfnamefont{D.}~\bibnamefont{Scanlan}},
  \bibinfo{author}{\bibfnamefont{N.}~\bibnamefont{McEvoy}},
  \bibinfo{author}{\bibfnamefont{G.~S.} \bibnamefont{Duesberg}},
  \bibinfo{author}{\bibfnamefont{G.~L.~W.} \bibnamefont{Cross}},
  \bibinfo{author}{\bibfnamefont{H.~Z.} \bibnamefont{Zhang}}, \bibnamefont{and}
  \bibinfo{author}{\bibfnamefont{J.~F.} \bibnamefont{Donegan}},
  \bibinfo{journal}{Applied Physics Letters} \textbf{\bibinfo{volume}{104}},
  \bibinfo{pages}{031109} (\bibinfo{year}{2014}), ISSN
  \bibinfo{issn}{0003-6951, 1077-3118},
  \urlprefix\url{http://aip.scitation.org/doi/10.1063/1.4862331}.

\bibitem[{\citenamefont{Tallaire et~al.}(2019)\citenamefont{Tallaire, Brinza,
  De~Feudis, Ferrier, Touati, Binet, Nicolas, Delord, Hétet, Herzig
  et~al.}}]{tallaire_synthesis_2019}
\bibinfo{author}{\bibfnamefont{A.}~\bibnamefont{Tallaire}},
  \bibinfo{author}{\bibfnamefont{O.}~\bibnamefont{Brinza}},
  \bibinfo{author}{\bibfnamefont{M.}~\bibnamefont{De~Feudis}},
  \bibinfo{author}{\bibfnamefont{A.}~\bibnamefont{Ferrier}},
  \bibinfo{author}{\bibfnamefont{N.}~\bibnamefont{Touati}},
  \bibinfo{author}{\bibfnamefont{L.}~\bibnamefont{Binet}},
  \bibinfo{author}{\bibfnamefont{L.}~\bibnamefont{Nicolas}},
  \bibinfo{author}{\bibfnamefont{T.}~\bibnamefont{Delord}},
  \bibinfo{author}{\bibfnamefont{G.}~\bibnamefont{Hétet}},
  \bibinfo{author}{\bibfnamefont{T.}~\bibnamefont{Herzig}},
  \bibnamefont{et~al.}, \bibinfo{journal}{ACS Applied Nano Materials}
  \textbf{\bibinfo{volume}{2}}, \bibinfo{pages}{5952} (\bibinfo{year}{2019}),
  ISSN \bibinfo{issn}{2574-0970, 2574-0970},
  \urlprefix\url{https://pubs.acs.org/doi/10.1021/acsanm.9b01395}.

\bibitem[{\citenamefont{Dolde et~al.}(2013)\citenamefont{Dolde, Jakobi,
  Naydenov, Zhao, Pezzagna, Trautmann, Meijer, Neumann, Jelezko, and
  Wrachtrup}}]{dolde_room-temperature_2013}
\bibinfo{author}{\bibfnamefont{F.}~\bibnamefont{Dolde}},
  \bibinfo{author}{\bibfnamefont{I.}~\bibnamefont{Jakobi}},
  \bibinfo{author}{\bibfnamefont{B.}~\bibnamefont{Naydenov}},
  \bibinfo{author}{\bibfnamefont{N.}~\bibnamefont{Zhao}},
  \bibinfo{author}{\bibfnamefont{S.}~\bibnamefont{Pezzagna}},
  \bibinfo{author}{\bibfnamefont{C.}~\bibnamefont{Trautmann}},
  \bibinfo{author}{\bibfnamefont{J.}~\bibnamefont{Meijer}},
  \bibinfo{author}{\bibfnamefont{P.}~\bibnamefont{Neumann}},
  \bibinfo{author}{\bibfnamefont{F.}~\bibnamefont{Jelezko}}, \bibnamefont{and}
  \bibinfo{author}{\bibfnamefont{J.}~\bibnamefont{Wrachtrup}},
  \bibinfo{journal}{Nature Physics} \textbf{\bibinfo{volume}{9}},
  \bibinfo{pages}{139} (\bibinfo{year}{2013}), ISSN \bibinfo{issn}{1745-2473,
  1745-2481}, \urlprefix\url{http://www.nature.com/articles/nphys2545}.

\bibitem[{\citenamefont{Acosta et~al.}(2009)\citenamefont{Acosta, Bauch,
  Ledbetter, Santori, Fu, Barclay, Beausoleil, Linget, Roch, Treussart
  et~al.}}]{acosta_diamonds_2009}
\bibinfo{author}{\bibfnamefont{V.~M.} \bibnamefont{Acosta}},
  \bibinfo{author}{\bibfnamefont{E.}~\bibnamefont{Bauch}},
  \bibinfo{author}{\bibfnamefont{M.~P.} \bibnamefont{Ledbetter}},
  \bibinfo{author}{\bibfnamefont{C.}~\bibnamefont{Santori}},
  \bibinfo{author}{\bibfnamefont{K.-M.~C.} \bibnamefont{Fu}},
  \bibinfo{author}{\bibfnamefont{P.~E.} \bibnamefont{Barclay}},
  \bibinfo{author}{\bibfnamefont{R.~G.} \bibnamefont{Beausoleil}},
  \bibinfo{author}{\bibfnamefont{H.}~\bibnamefont{Linget}},
  \bibinfo{author}{\bibfnamefont{J.~F.} \bibnamefont{Roch}},
  \bibinfo{author}{\bibfnamefont{F.}~\bibnamefont{Treussart}},
  \bibnamefont{et~al.}, \bibinfo{journal}{Physical Review B}
  \textbf{\bibinfo{volume}{80}}, \bibinfo{pages}{115202}
  (\bibinfo{year}{2009}), ISSN \bibinfo{issn}{1098-0121, 1550-235X},
  \urlprefix\url{https://link.aps.org/doi/10.1103/PhysRevB.80.115202}.

\bibitem[{\citenamefont{Himics et~al.}(2015)\citenamefont{Himics, Tóth, Veres,
  Tóth, and Koós}}]{himics_effective_2015}
\bibinfo{author}{\bibfnamefont{L.}~\bibnamefont{Himics}},
  \bibinfo{author}{\bibfnamefont{S.}~\bibnamefont{Tóth}},
  \bibinfo{author}{\bibfnamefont{M.}~\bibnamefont{Veres}},
  \bibinfo{author}{\bibfnamefont{A.}~\bibnamefont{Tóth}}, \bibnamefont{and}
  \bibinfo{author}{\bibfnamefont{M.}~\bibnamefont{Koós}},
  \bibinfo{journal}{Applied Surface Science} \textbf{\bibinfo{volume}{328}},
  \bibinfo{pages}{577} (\bibinfo{year}{2015}), ISSN \bibinfo{issn}{01694332},
  \urlprefix\url{https://linkinghub.elsevier.com/retrieve/pii/S0169433214028025}.

\bibitem[{\citenamefont{Ziegler et~al.}(2010)\citenamefont{Ziegler, Ziegler,
  and Biersack}}]{ziegler_srim_2010}
\bibinfo{author}{\bibfnamefont{J.~F.} \bibnamefont{Ziegler}},
  \bibinfo{author}{\bibfnamefont{M.~D.} \bibnamefont{Ziegler}},
  \bibnamefont{and} \bibinfo{author}{\bibfnamefont{J.~P.}
  \bibnamefont{Biersack}}, \bibinfo{journal}{Nuclear Instruments and Methods in
  Physics Research Section B: Beam Interactions with Materials and Atoms}
  \textbf{\bibinfo{volume}{268}}, \bibinfo{pages}{1818} (\bibinfo{year}{2010}),
  ISSN \bibinfo{issn}{0168-583X},
  \urlprefix\url{http://www.sciencedirect.com/science/article/pii/S0168583X10001862}.

\bibitem[{\citenamefont{Forneris et~al.}(2016)\citenamefont{Forneris,
  Tengattini, Tchernij, Picollo, Battiato, Traina, Degiovanni, Moreva, Brida,
  Grilj et~al.}}]{forneris_creation_2016}
\bibinfo{author}{\bibfnamefont{J.}~\bibnamefont{Forneris}},
  \bibinfo{author}{\bibfnamefont{A.}~\bibnamefont{Tengattini}},
  \bibinfo{author}{\bibfnamefont{S.~D.} \bibnamefont{Tchernij}},
  \bibinfo{author}{\bibfnamefont{F.}~\bibnamefont{Picollo}},
  \bibinfo{author}{\bibfnamefont{A.}~\bibnamefont{Battiato}},
  \bibinfo{author}{\bibfnamefont{P.}~\bibnamefont{Traina}},
  \bibinfo{author}{\bibfnamefont{I.}~\bibnamefont{Degiovanni}},
  \bibinfo{author}{\bibfnamefont{E.}~\bibnamefont{Moreva}},
  \bibinfo{author}{\bibfnamefont{G.}~\bibnamefont{Brida}},
  \bibinfo{author}{\bibfnamefont{V.}~\bibnamefont{Grilj}},
  \bibnamefont{et~al.}, \bibinfo{journal}{Journal of Luminescence}
  \textbf{\bibinfo{volume}{179}}, \bibinfo{pages}{59} (\bibinfo{year}{2016}),
  ISSN \bibinfo{issn}{00222313},
  \urlprefix\url{https://linkinghub.elsevier.com/retrieve/pii/S0022231316306275}.

\bibitem[{\citenamefont{Bauch et~al.}(2018)\citenamefont{Bauch, Hart, Schloss,
  Turner, Barry, Kehayias, Singh, and Walsworth}}]{bauch_ultralong_2018}
\bibinfo{author}{\bibfnamefont{E.}~\bibnamefont{Bauch}},
  \bibinfo{author}{\bibfnamefont{C.~A.} \bibnamefont{Hart}},
  \bibinfo{author}{\bibfnamefont{J.~M.} \bibnamefont{Schloss}},
  \bibinfo{author}{\bibfnamefont{M.~J.} \bibnamefont{Turner}},
  \bibinfo{author}{\bibfnamefont{J.~F.} \bibnamefont{Barry}},
  \bibinfo{author}{\bibfnamefont{P.}~\bibnamefont{Kehayias}},
  \bibinfo{author}{\bibfnamefont{S.}~\bibnamefont{Singh}}, \bibnamefont{and}
  \bibinfo{author}{\bibfnamefont{R.~L.} \bibnamefont{Walsworth}},
  \bibinfo{journal}{Physical Review X} \textbf{\bibinfo{volume}{8}},
  \bibinfo{pages}{031025} (\bibinfo{year}{2018}), ISSN
  \bibinfo{issn}{2160-3308},
  \urlprefix\url{https://link.aps.org/doi/10.1103/PhysRevX.8.031025}.

\bibitem[{\citenamefont{Dréau et~al.}(2012)\citenamefont{Dréau, Maze, Lesik,
  Roch, and Jacques}}]{dreau_high-resolution_2012}
\bibinfo{author}{\bibfnamefont{A.}~\bibnamefont{Dréau}},
  \bibinfo{author}{\bibfnamefont{J.-R.} \bibnamefont{Maze}},
  \bibinfo{author}{\bibfnamefont{M.}~\bibnamefont{Lesik}},
  \bibinfo{author}{\bibfnamefont{J.-F.} \bibnamefont{Roch}}, \bibnamefont{and}
  \bibinfo{author}{\bibfnamefont{V.}~\bibnamefont{Jacques}},
  \bibinfo{journal}{Physical Review B} \textbf{\bibinfo{volume}{85}},
  \bibinfo{pages}{134107} (\bibinfo{year}{2012}), ISSN
  \bibinfo{issn}{1098-0121, 1550-235X},
  \urlprefix\url{https://link.aps.org/doi/10.1103/PhysRevB.85.134107}.

\bibitem[{\citenamefont{Mizuochi et~al.}(2009)\citenamefont{Mizuochi, Neumann,
  Rempp, Beck, Jacques, Siyushev, Nakamura, Twitchen, Watanabe, Yamasaki
  et~al.}}]{mizuochi_coherence_2009}
\bibinfo{author}{\bibfnamefont{N.}~\bibnamefont{Mizuochi}},
  \bibinfo{author}{\bibfnamefont{P.}~\bibnamefont{Neumann}},
  \bibinfo{author}{\bibfnamefont{F.}~\bibnamefont{Rempp}},
  \bibinfo{author}{\bibfnamefont{J.}~\bibnamefont{Beck}},
  \bibinfo{author}{\bibfnamefont{V.}~\bibnamefont{Jacques}},
  \bibinfo{author}{\bibfnamefont{P.}~\bibnamefont{Siyushev}},
  \bibinfo{author}{\bibfnamefont{K.}~\bibnamefont{Nakamura}},
  \bibinfo{author}{\bibfnamefont{D.~J.} \bibnamefont{Twitchen}},
  \bibinfo{author}{\bibfnamefont{H.}~\bibnamefont{Watanabe}},
  \bibinfo{author}{\bibfnamefont{S.}~\bibnamefont{Yamasaki}},
  \bibnamefont{et~al.}, \bibinfo{journal}{Physical Review B}
  \textbf{\bibinfo{volume}{80}}, \bibinfo{pages}{041201}
  (\bibinfo{year}{2009}), ISSN \bibinfo{issn}{1098-0121, 1550-235X},
  \urlprefix\url{https://link.aps.org/doi/10.1103/PhysRevB.80.041201}.

\bibitem[{\citenamefont{Jamonneau et~al.}(2016)\citenamefont{Jamonneau, Lesik,
  Tetienne, Alvizu, Mayer, Dréau, Kosen, Roch, Pezzagna, Meijer
  et~al.}}]{jamonneau_competition_2016}
\bibinfo{author}{\bibfnamefont{P.}~\bibnamefont{Jamonneau}},
  \bibinfo{author}{\bibfnamefont{M.}~\bibnamefont{Lesik}},
  \bibinfo{author}{\bibfnamefont{J.~P.} \bibnamefont{Tetienne}},
  \bibinfo{author}{\bibfnamefont{I.}~\bibnamefont{Alvizu}},
  \bibinfo{author}{\bibfnamefont{L.}~\bibnamefont{Mayer}},
  \bibinfo{author}{\bibfnamefont{A.}~\bibnamefont{Dréau}},
  \bibinfo{author}{\bibfnamefont{S.}~\bibnamefont{Kosen}},
  \bibinfo{author}{\bibfnamefont{J.-F.} \bibnamefont{Roch}},
  \bibinfo{author}{\bibfnamefont{S.}~\bibnamefont{Pezzagna}},
  \bibinfo{author}{\bibfnamefont{J.}~\bibnamefont{Meijer}},
  \bibnamefont{et~al.}, \bibinfo{journal}{Physical Review B}
  \textbf{\bibinfo{volume}{93}}, \bibinfo{pages}{024305}
  (\bibinfo{year}{2016}), ISSN \bibinfo{issn}{2469-9950, 2469-9969},
  \urlprefix\url{https://link.aps.org/doi/10.1103/PhysRevB.93.024305}.

\bibitem[{\citenamefont{Acosta et~al.}(2010)\citenamefont{Acosta, Bauch,
  Ledbetter, Waxman, Bouchard, and Budker}}]{acosta_temperature_2010}
\bibinfo{author}{\bibfnamefont{V.~M.} \bibnamefont{Acosta}},
  \bibinfo{author}{\bibfnamefont{E.}~\bibnamefont{Bauch}},
  \bibinfo{author}{\bibfnamefont{M.~P.} \bibnamefont{Ledbetter}},
  \bibinfo{author}{\bibfnamefont{A.}~\bibnamefont{Waxman}},
  \bibinfo{author}{\bibfnamefont{L.-S.} \bibnamefont{Bouchard}},
  \bibnamefont{and} \bibinfo{author}{\bibfnamefont{D.}~\bibnamefont{Budker}},
  \bibinfo{journal}{Physical Review Letters} \textbf{\bibinfo{volume}{104}},
  \bibinfo{pages}{070801} (\bibinfo{year}{2010}), ISSN
  \bibinfo{issn}{0031-9007, 1079-7114},
  \urlprefix\url{https://link.aps.org/doi/10.1103/PhysRevLett.104.070801}.

\bibitem[{\citenamefont{Bauch et~al.}(2020)\citenamefont{Bauch, Singh, Lee,
  Hart, Schloss, Turner, Barry, Pham, Bar-Gill, Yelin
  et~al.}}]{bauch_decoherence_2020}
\bibinfo{author}{\bibfnamefont{E.}~\bibnamefont{Bauch}},
  \bibinfo{author}{\bibfnamefont{S.}~\bibnamefont{Singh}},
  \bibinfo{author}{\bibfnamefont{J.}~\bibnamefont{Lee}},
  \bibinfo{author}{\bibfnamefont{C.~A.} \bibnamefont{Hart}},
  \bibinfo{author}{\bibfnamefont{J.~M.} \bibnamefont{Schloss}},
  \bibinfo{author}{\bibfnamefont{M.~J.} \bibnamefont{Turner}},
  \bibinfo{author}{\bibfnamefont{J.~F.} \bibnamefont{Barry}},
  \bibinfo{author}{\bibfnamefont{L.~M.} \bibnamefont{Pham}},
  \bibinfo{author}{\bibfnamefont{N.}~\bibnamefont{Bar-Gill}},
  \bibinfo{author}{\bibfnamefont{S.~F.} \bibnamefont{Yelin}},
  \bibnamefont{et~al.}, \bibinfo{journal}{Physical Review B}
  \textbf{\bibinfo{volume}{102}}, \bibinfo{pages}{134210}
  (\bibinfo{year}{2020}), ISSN \bibinfo{issn}{2469-9950, 2469-9969},
  \urlprefix\url{https://link.aps.org/doi/10.1103/PhysRevB.102.134210}.

\bibitem[{\citenamefont{Jarmola et~al.}(2012)\citenamefont{Jarmola, Acosta,
  Jensen, Chemerisov, and Budker}}]{jarmola_temperature-_2012}
\bibinfo{author}{\bibfnamefont{A.}~\bibnamefont{Jarmola}},
  \bibinfo{author}{\bibfnamefont{V.~M.} \bibnamefont{Acosta}},
  \bibinfo{author}{\bibfnamefont{K.}~\bibnamefont{Jensen}},
  \bibinfo{author}{\bibfnamefont{S.}~\bibnamefont{Chemerisov}},
  \bibnamefont{and} \bibinfo{author}{\bibfnamefont{D.}~\bibnamefont{Budker}},
  \bibinfo{journal}{Physical Review Letters} \textbf{\bibinfo{volume}{108}},
  \bibinfo{pages}{197601} (\bibinfo{year}{2012}), ISSN
  \bibinfo{issn}{0031-9007, 1079-7114},
  \urlprefix\url{https://link.aps.org/doi/10.1103/PhysRevLett.108.197601}.

\bibitem[{\citenamefont{Farfurnik et~al.}(2017)\citenamefont{Farfurnik, Alfasi,
  Masis, Kauffmann, Farchi, Romach, Hovav, Buks, and
  Bar-Gill}}]{farfurnik_enhanced_2017}
\bibinfo{author}{\bibfnamefont{D.}~\bibnamefont{Farfurnik}},
  \bibinfo{author}{\bibfnamefont{N.}~\bibnamefont{Alfasi}},
  \bibinfo{author}{\bibfnamefont{S.}~\bibnamefont{Masis}},
  \bibinfo{author}{\bibfnamefont{Y.}~\bibnamefont{Kauffmann}},
  \bibinfo{author}{\bibfnamefont{E.}~\bibnamefont{Farchi}},
  \bibinfo{author}{\bibfnamefont{Y.}~\bibnamefont{Romach}},
  \bibinfo{author}{\bibfnamefont{Y.}~\bibnamefont{Hovav}},
  \bibinfo{author}{\bibfnamefont{E.}~\bibnamefont{Buks}}, \bibnamefont{and}
  \bibinfo{author}{\bibfnamefont{N.}~\bibnamefont{Bar-Gill}},
  \bibinfo{journal}{Applied Physics Letters} \textbf{\bibinfo{volume}{111}},
  \bibinfo{pages}{123101} (\bibinfo{year}{2017}), ISSN
  \bibinfo{issn}{0003-6951, 1077-3118},
  \urlprefix\url{http://aip.scitation.org/doi/10.1063/1.4993257}.

\bibitem[{\citenamefont{Ishikawa et~al.}(2012)\citenamefont{Ishikawa, Fu,
  Santori, Acosta, Beausoleil, Watanabe, Shikata, and
  Itoh}}]{ishikawa_optical_2012}
\bibinfo{author}{\bibfnamefont{T.}~\bibnamefont{Ishikawa}},
  \bibinfo{author}{\bibfnamefont{K.-M.~C.} \bibnamefont{Fu}},
  \bibinfo{author}{\bibfnamefont{C.}~\bibnamefont{Santori}},
  \bibinfo{author}{\bibfnamefont{V.~M.} \bibnamefont{Acosta}},
  \bibinfo{author}{\bibfnamefont{R.~G.} \bibnamefont{Beausoleil}},
  \bibinfo{author}{\bibfnamefont{H.}~\bibnamefont{Watanabe}},
  \bibinfo{author}{\bibfnamefont{S.}~\bibnamefont{Shikata}}, \bibnamefont{and}
  \bibinfo{author}{\bibfnamefont{K.~M.} \bibnamefont{Itoh}},
  \bibinfo{journal}{Nano Letters} \textbf{\bibinfo{volume}{12}},
  \bibinfo{pages}{2083} (\bibinfo{year}{2012}), ISSN \bibinfo{issn}{1530-6984,
  1530-6992}, \urlprefix\url{https://pubs.acs.org/doi/10.1021/nl300350r}.

\bibitem[{\citenamefont{Deák et~al.}(2014)\citenamefont{Deák, Aradi, Kaviani,
  Frauenheim, and Gali}}]{deak_formation_2014}
\bibinfo{author}{\bibfnamefont{P.}~\bibnamefont{Deák}},
  \bibinfo{author}{\bibfnamefont{B.}~\bibnamefont{Aradi}},
  \bibinfo{author}{\bibfnamefont{M.}~\bibnamefont{Kaviani}},
  \bibinfo{author}{\bibfnamefont{T.}~\bibnamefont{Frauenheim}},
  \bibnamefont{and} \bibinfo{author}{\bibfnamefont{A.}~\bibnamefont{Gali}},
  \bibinfo{journal}{Physical Review B} \textbf{\bibinfo{volume}{89}},
  \bibinfo{pages}{075203} (\bibinfo{year}{2014}), ISSN
  \bibinfo{issn}{1098-0121, 1550-235X},
  \urlprefix\url{https://link.aps.org/doi/10.1103/PhysRevB.89.075203}.

\bibitem[{\citenamefont{Bar-Gill et~al.}(2013)\citenamefont{Bar-Gill, Pham,
  Jarmola, Budker, and Walsworth}}]{bar-gill_solid-state_2013}
\bibinfo{author}{\bibfnamefont{N.}~\bibnamefont{Bar-Gill}},
  \bibinfo{author}{\bibfnamefont{L.}~\bibnamefont{Pham}},
  \bibinfo{author}{\bibfnamefont{A.}~\bibnamefont{Jarmola}},
  \bibinfo{author}{\bibfnamefont{D.}~\bibnamefont{Budker}}, \bibnamefont{and}
  \bibinfo{author}{\bibfnamefont{R.}~\bibnamefont{Walsworth}},
  \bibinfo{journal}{Nature Communications} \textbf{\bibinfo{volume}{4}},
  \bibinfo{pages}{1743} (\bibinfo{year}{2013}), ISSN \bibinfo{issn}{2041-1723},
  \urlprefix\url{http://www.nature.com/articles/ncomms2771}.

\end{thebibliography}

\end{document}